\documentclass[preprint,12pt]{elsarticle}
\usepackage[a4paper, margin=2.5cm]{geometry}

\usepackage{xcolor} 
\definecolor{airforceblue}{rgb}{0.36, 0.54, 0.66}
\definecolor{electriclime}{rgb}{0.8, 1.0, 0.0}
\definecolor{darkgreen}{rgb}{0.0, 0.5, 0.0}
\definecolor{codegreen}{rgb}{0,0.6,0}
\definecolor{codegray}{rgb}{0.5,0.5,0.5}
\definecolor{codepurple}{rgb}{0.58,0,0.82}
\definecolor{backcolour}{rgb}{0.95,0.95,0.92}

\usepackage{graphicx}
\usepackage{float}
\graphicspath{{figures/}}

\usepackage{outlines}
\usepackage{realboxes}
\usepackage[utf8]{inputenc}
\usepackage{textcomp}
\usepackage{caption}
\usepackage{subcaption}
\usepackage{amsmath}

\usepackage{listings}
\lstdefinestyle{mystyle}{
    backgroundcolor=\color{backcolour},   
    commentstyle=\color{codegreen},
    keywordstyle=\color{magenta},
    numberstyle=\tiny\color{codegray},
    stringstyle=\color{codepurple},
    basicstyle=\ttfamily\footnotesize,
    breakatwhitespace=false,         
    breaklines=true,                 
    captionpos=b,                    
    keepspaces=true,                 
    numbers=left,                    
    numbersep=5pt,                  
    showspaces=false,                
    showstringspaces=false,
    showtabs=false,                  
    tabsize=2
}
\usepackage{amssymb}
\usepackage{upgreek}
\usepackage{gensymb}

\usepackage{verbatim}
\usepackage{alltt}

\usepackage{rotating}
\usepackage{subfiles} 

\makeatletter
\def\ps@pprintTitle{%
  \let\@oddhead\@empty
  \let\@evenhead\@empty
  \let\@oddfoot\@empty
  \let\@evenfoot\@oddfoot
}
\makeatother

\usepackage{hyperref}
\hypersetup{
    colorlinks=true,
    linkcolor=blue,
    filecolor=magenta,      
    urlcolor=cyan,
}
\begin{document}

\begin{frontmatter}

\title{gSeaGen code by KM3NeT: an efficient tool to propagate muons simulated with CORSIKA}

\cortext[cor]{corresponding author}

\author[a]{S.~Aiello}
\author[b,bb]{A.~Albert}
\author[c]{A.\,R.~Alhebsi}
\author[d]{M.~Alshamsi}
\author[e]{S. Alves Garre}
\author[g,f]{A. Ambrosone}
\author[h]{F.~Ameli}
\author[i]{M.~Andre}
\author[j]{L.~Aphecetche}
\author[k]{M. Ardid}
% ...
\address[a]{INFN, Sezione di Catania, (INFN-CT) Via Santa Sofia 64, Catania, 95123 Italy}
\address[b]{Universit{\'e}~de~Strasbourg,~CNRS,~IPHC~UMR~7178,~F-67000~Strasbourg,~France}
\address[c]{Khalifa University, Department of Physics, PO Box 127788, Abu Dhabi, 0000 United Arab Emirates}
\address[d]{Aix~Marseille~Univ,~CNRS/IN2P3,~CPPM,~Marseille,~France}
\address[e]{IFIC - Instituto de F{\'\i}sica Corpuscular (CSIC - Universitat de Val{\`e}ncia), c/Catedr{\'a}tico Jos{\'e} Beltr{\'a}n, 2, 46980 Paterna, Valencia, Spain}
\address[f]{INFN, Sezione di Napoli, Complesso Universitario di Monte S. Angelo, Via Cintia ed. G, Napoli, 80126 Italy}
\address[g]{Universit{\`a} di Napoli ``Federico II'', Dip. Scienze Fisiche ``E. Pancini'', Complesso Universitario di Monte S. Angelo, Via Cintia ed. G, Napoli, 80126 Italy}
\address[h]{INFN, Sezione di Roma, Piazzale Aldo Moro 2, Roma, 00185 Italy}
\address[i]{Universitat Polit{\`e}cnica de Catalunya, Laboratori d'Aplicacions Bioac{\'u}stiques, Centre Tecnol{\`o}gic de Vilanova i la Geltr{\'u}, Avda. Rambla Exposici{\'o}, s/n, Vilanova i la Geltr{\'u}, 08800 Spain}
\address[j]{Subatech, IMT Atlantique, IN2P3-CNRS, Nantes Universit{\'e}, 4 rue Alfred Kastler - La Chantrerie, Nantes, BP 20722 44307 France}
\address[k]{Universitat Polit{\`e}cnica de Val{\`e}ncia, Instituto de Investigaci{\'o}n para la Gesti{\'o}n Integrada de las Zonas Costeras, C/ Paranimf, 1, Gandia, 46730 Spain}
% ...
\address[bb]{Universit{\'e} de Haute Alsace, rue des Fr{\`e}res Lumi{\`e}re, 68093 Mulhouse Cedex, France}
% ...

\begin{abstract}
    {The KM3NeT Collaboration has tackled a common challenge faced by the astroparticle physics community, namely adapting the experiment-specific simulation software to work with the CORSIKA air shower simulation output. The proposed solution is an extension of the open source code gSeaGen, which allows the transport of muons generated by CORSIKA to a detector of any size at an arbitrary depth. The gSeaGen code was not only extended in terms of functionality but also underwent a thorough redesign of the muon propagation routine, resulting in a more accurate and efficient simulation. This paper presents the capabilities of the new gSeaGen code as well as prospects for further developments.}
\end{abstract}
\label{abstract}

\end{frontmatter}




\section{Introduction}
\label{sec:intro}


Despite several groundbreaking observations in recent years \cite{IceCube_blazar,IceCube_blazar_multimessenger,aiello_observation_2025,GW170817_first_GW_from_binary_neutron_star,KM3NeT_GW_counterparts,repeating_FRB,GW190521_binary_BH_merger,M87_BH_image}, current understanding of the universe at extreme energies is still limited due to the scarcity of available data. The limiting factors are the low fluxes of high-energy particles and difficulties in tying detections to particular sources, due to interactions with interstellar matter and bending by the magnetic fields along their way \cite{UHE-CR-facts-and-myths,Magnetic_fields_and_CRs}. Neutrinos are the best known solution to the latter problem; electrically neutral and interacting through weak force, they escape unbowed, unbent, and unbroken even from the most dense environments \cite{IceCube_blazar,IceCube_blazar_multimessenger,Borexino_solar_MSW,CCSN_KM3NeT_2021}. Low interaction cross sections give neutrinos an advantage over other particles in terms of escaping their region of production, but at the same time they make it a challenge to detect them \cite{Neutrino_cross-sections}. Thus, to measure even atmospheric neutrinos, let alone cosmic ones, one must first deal with the overwhelming background of atmospheric muons. The most straightforward solution is to move the detectors deeper, where the muons are sufficiently attenuated, although even at several km w.e. they still constitute a non-negligible signal \cite{Andrey_and_me_ICRC2023,ARCA6_results,IceCube_seasonal_muons,DepthDependenceMassimiliano}. To develop efficient cuts and obtain an atmospheric or astrophysical neutrino sample, a detailed simulation of the expected muon flux is required \cite{Aiello2019,ORCA_oscillations}.

Since the introduction of the CORSIKA air shower simulation software \cite{CORSIKA}, various underground and underwater experiments have invested efforts in transporting the output particles from the ground / sea level to their detectors. The KM3NeT Collaboration utilises a dedicated software, gSeaGen, to solve the problem of transporting particles produced with the GENIE MC generator \cite{GENIE} from the interaction vertex to the detector. gSeaGen is an open-source software that can be accessed from the public GitLab repository \cite{gSeaGen_repo} or the Zenodo record \cite{gSeaGen_Zenodo}. Its features are explained in more detail in \cite{gSeaGen-2020}. This paper describes the modifications made to gSeaGen to read the CORSIKA binary output files and track potentially interesting particles, namely high-energy atmospheric muons\footnote{Currently, atmospheric neutrinos and other secondary particles simulated by CORSIKA are not transported.}, from sea level to deep sea. The new functionalities are of high importance for the KM3NeT neutrino telescopes: KM3NeT/ARCA, dedicated to the detection of high-energy neutrinos from astrophysical sources, and KM3NeT/ORCA, dedicated to the measurement of neutrino mass ordering and oscillation parameters with atmospheric neutrinos \cite{KM3NeT-LoI-2.0}. The new features and improvements are presented in the following sections.

\section{Detector positioning}
\label{sec:coordinate_system}

\begin{figure}[H]
\centering
\includegraphics[scale=0.5]{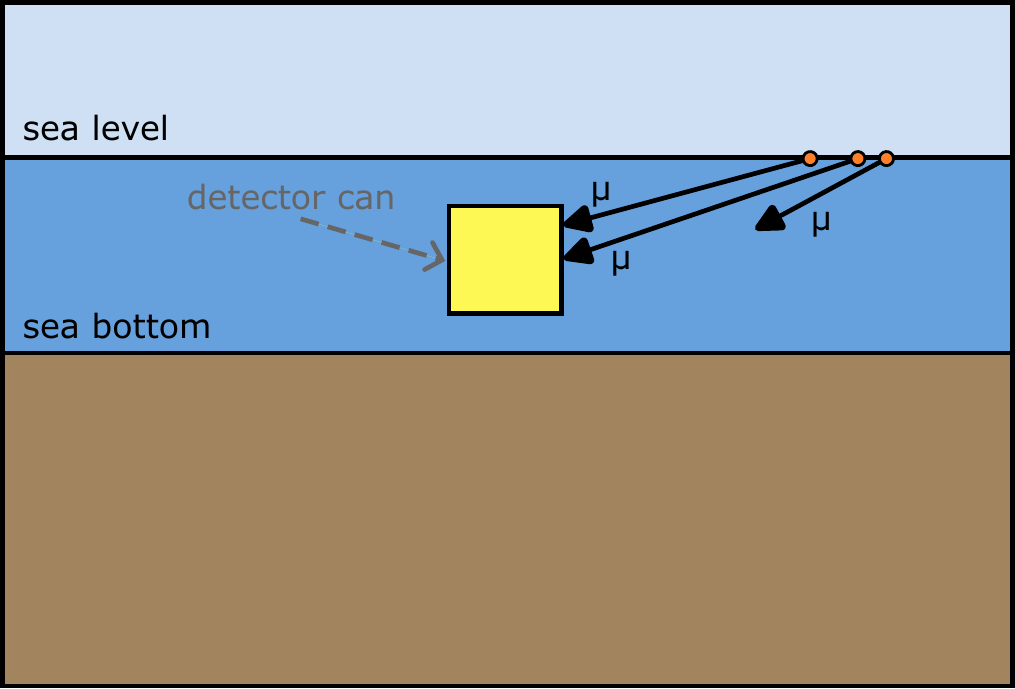}
\caption
{Sketch of the new geometry in gSeaGen, with an example of a multi-muon CORSIKA event, with two muons reaching the can and one stopping in seawater. The curvature of the Earth has been neglected in this drawing, however it is explicitly handled by the gSeaGen code, as detailed in Sec. \ref{sec:rotation}}

\label{fig:new_geometry2}       
\end{figure}

The KM3NeT neutrino telescopes are arrays of optical sensors installed deep under the sea surface. Their geometrical layout was optimised according to their main scientific goals. For simulation purposes, the sensitive volume of each telescope can be defined as a virtual cylindrical region around the strings carrying the optical sensors, called the “can”. The concept of a “can” in the context of underwater neutrino telescopes is explained in more detail in \cite{ANTARES_Monte_Carlo_with_can_description,MUPAGE}. The size of the can is defined by its height and radius. The user can adjust both parameters, which, depending on the context, may take into account the lateral distance of muons in a bundle or the optical properties of the deep-sea site. 
The main novelty introduced with the modified gSeaGen code is the possibility of reading and tracking CORSIKA muons from sea level to can. Moreover, the bottom of the can is no longer fixed on the seabed and can be placed in a different position (see Fig.~\ref{fig:new_geometry2}). This feature makes the code more flexible and allows for a more accurate simulation of the detector geometry. 

The media for particle propagation in gSeaGen are not limited to seawater and rock, which are the KM3NeT defaults. Different types of rock, ice and water with variable optical properties can be used, as indicated in \cite{gSeaGen-2020}. This allows for the adaptation of the gSeaGen geometry to several existing or planned experiments, in particular: IceCube \cite{IceCube-basic-paper}, ANTARES \cite{ANTARES}, Baikal-GVD \cite{Baikal-GVD}, Super-Kamiokande \cite{SuperKamiokande}, Hyper-Kamiokande \cite{Hyper_Kamiokande}, P-ONE \cite{P-ONE}, TRIDENT \cite{TRIDENT}, HUNT \cite{HUNT} and NEON \cite{NEON}.

\section{Processing of CORSIKA}
\label{sec:BIN_processing}

The command to pass a CORSIKA binary file to gSeaGen is:

\begin{lstlisting}[language=bash]
gSeaNuEvGen -f "BIN:$DIR/DAT$RUN"
\end{lstlisting}

This invokes a dedicated CORSIKA file driver \emph{GSeaCORSIKAFileFlux.cxx}, developed from the \emph{readcorsika.cpp} script, distributed together with CORSIKA. The implementation was based on and was tested primarily with CORSIKA v7.7410. Backward compatibility is mostly bound to the changes in the output format of CORSIKA, however, the earliest tested version was v7.4005.

\subsection{Track information}
\label{sec:track-info}

\hfill

The properties of the particles are stored in track objects in gSeaGen. To preserve the mother-daughter relation between the particles in the gSeaGen processing, three track categories have been introduced:
\begin{enumerate}
    \item Mother --- mother particle, which created the muon by decaying or interacting.
    \item Grandmother --- grandmother particle, which created the mother particle by decaying or interacting.
    \item Muaddi --- muon at the production point.
\end{enumerate}

The hadronic and electromagnetic (EM) generation counters available from CORSIKA are saved as follows:
\begin{itemize}
  \item grandmother track: EM generation counter of the mother particle,
  \item mother track: hadronic generation counter of the mother particle,
  \item muaddi track: hadronic generation counter of the muon,
  \item muon track: extended EM generation counter of the muon.
\end{itemize}

\begin{table}[H]
\caption{
Summary of the information stored for each of the track categories. The definition of each category is given in the text. {\fontfamily{qcr}\selectfont N/A} represents information which is not available from CORSIKA. In the table, only muon tracks are indicated on the bottom, since currently gSeaGen only propagates muons from CORSIKA. Propagation of other secondary particles produced by CORSIKA is not yet supported. {\fontfamily{qcr}\selectfont ID} is the particle identification number converted to the PDG numbering scheme \cite{PDG}, $t$ is the time and $l$ is the travelled distance. The abbreviation {\fontfamily{qcr}\selectfont obslev} stands for the observation level (specified in CORSIKA), which in the case of the KM3NeT experiment is the sea level. The {\fontfamily{qcr}\selectfont status} field is described in Sec.\ \ref{sec:status_values} and {\fontfamily{qcr}\selectfont counter} is the hadronic or electromagnetic interaction counter, used in CORSIKA.
}
\centering
\resizebox{\columnwidth}{!}{

\begin{tabular}{|c||c|c|c|c|c|c|c|c|c|c|c|c|c|}
\hline 
{Track} & {$x$} & {$y$} & {$z$} & {$\frac{p_{x}}{p}$} & {$\frac{p_{y}}{p}$} & {$\frac{p_{z}}{p}$} & {$E$} & {$t$} & {ID} & {$l$} & {$E_{\mathrm{loss}}$} & {status} & {counter}\tabularnewline
\hline 
\hline 
{primary} & {$0$} & {$0$} & {$\checkmark$} & {$\checkmark$} & {$\checkmark$} & {$\checkmark$} & {$\checkmark$} & {$0$} & {$\checkmark$} & {$0$} & {N/A} & {200}& {N/A}\tabularnewline
\hline 
{muaddi} & {$\checkmark$} & {$\checkmark$} & {$\checkmark$} & {$\checkmark$} & {$\checkmark$} & {$\checkmark$} & {$\checkmark$} & {N/A} & {$\pm13$} & {N/A} & {N/A} & {-21}& {$\checkmark$}\tabularnewline
\hline 
{grandmother} & {$\checkmark$} & {$\checkmark$} & {$\checkmark$} & {$\checkmark$} & {$\checkmark$} & {$\checkmark$} & {$\checkmark$} & {N/A} & {$\checkmark$} & {N/A} & {N/A} & {\checkmark}& {$\checkmark$}\tabularnewline
\hline 
{mother} & {$\checkmark$} & {$\checkmark$} & {$\checkmark$} & {$\checkmark$} & {$\checkmark$} & {$\checkmark$} & {$\checkmark$} & {N/A} & {$\checkmark$} & {N/A} & {N/A} & {\checkmark}& {$\checkmark$}\tabularnewline
\hline 
{muon} & {$\checkmark$} & {$\checkmark$} & {=obslev} & {$\checkmark$} & {$\checkmark$} & {$\checkmark$} & {$\checkmark$} & {$\checkmark$} & {$\pm13$} & {$\checkmark$} & {$\checkmark$} & {\checkmark}& {$\checkmark$}\tabularnewline
\hline 
\end{tabular}{\tiny\par}%
}
{\small{}\vspace{0.2cm}
}{\small\par}
\label{tab:tracks-summary}
\end{table}

For further details on how the counters relate to each other and how they are incremented, see \cite{CORSIKA-Userguide}.

As reported in Tab.~\ref{tab:tracks-summary}, each track has a corresponding ID. Most of the particle IDs are simply converted from CORSIKA to the PDG numbering scheme \cite{PDG}. There are several exceptions where the CORSIKA ID has no PDG counterpart. In such cases, a related PDG ID is used and a special {\fontfamily{qcr}\selectfont status} value is assigned to avoid ambiguity, as shown in Tab.~\ref{tab:corsika-specific-pIDs}.

\begin{table}[H]
\centering
\caption{Summary of CORSIKA-specific IDs and their assigned PDG IDs and statuses.}
\begin{tabular}{|c||c|c|c|}
\hline 
{CORSIKA ID} & {description} & {PDG ID} & {status} \tabularnewline
\hline 
\hline 
{71} & {$\eta \rightarrow 2 \gamma$} & {221} & {24}\tabularnewline
\hline 
{72} & {$\eta \rightarrow 3 \pi^0$} & {221} & {25}\tabularnewline
\hline 
{73} & {$\eta \rightarrow \pi^+ \pi^- \pi^0$} & {221} & {26}\tabularnewline
\hline 
{74} & {$\eta \rightarrow \pi^+ \pi^- \gamma$} & {221} & {27}\tabularnewline
\hline 
{75} & {$\mu^+$ add. info} & {-13} & {24}\tabularnewline
\hline 
{76} & {$\mu^-$ add. info} & {13} & {24}\tabularnewline
\hline 
{85} & {decaying $\mu^+$ at start} & {-13} & {22}\tabularnewline
\hline 
{86} & {decaying $\mu^-$ at start} & {13} & {22}\tabularnewline
\hline 
{95} & {decaying $\mu^+$ at end} & {-13} & {23}\tabularnewline
\hline 
{96} & {decaying $\mu^-$ at end} & {13} & {23}\tabularnewline
\hline 
{9900} & {Cherenkov $\gamma$ on particle output file} & {22} & {28}\tabularnewline
\hline 
\end{tabular}{\tiny\par}

{\small{}\vspace{0.2cm}
}{\small\par}
\label{tab:corsika-specific-pIDs}
\end{table}

The CORSIKA output at sea level (the input for gSeaGen) contains more particles besides muons. Those are mainly neutrinos with a small fraction (approximately 3\% in the case of KM3NeT simulations at sea level) of other particles, like $\Sigma^-$, $\Lambda$, $p$, $n$, $K^{\pm}$, $\pi^{\pm}$, $K^0_L$, or $K^0_S$. The gSeaGen output may preserve the particle showers produced by CORSIKA at sea level and link them to propagated muons at the can level. Currently, non-muon particles can only be stored at sea level but not propagated. Not all muons at sea level undergo propagation. They are preselected, based on their energy and direction, which translates into the range and minimal distance they would have to travel to reach the can. This leads to three possible categories of muon tracks: unpropagated, propagated but not reaching the can, and propagated arriving at the can. More details may be found in Sec.\ \ref{sec:propagation}. The amount of saved information can be controlled using a combination of \Colorbox{backcolour}{\lstinline|-write|} and \Colorbox{backcolour}{\lstinline|-save|} options:

\begin{outline}[enumerate]
 \1 \Colorbox{backcolour}{\lstinline|-write 0|}: no mother particle tracks are saved except for the primaries, which started the simulated showers (events). The muons that underwent propagation and reached the can or belong to a shower reaching the can are stored in the output. If a single muon in a multi-muon event reaches the can, all muons in the event are saved.
 \1 \Colorbox{backcolour}{\lstinline|-write 1|}: the same muon tracks as with option \Colorbox{backcolour}{\lstinline|-write 0|} are saved, along with the tracks corresponding to their mother particles and with information about the muon creation point (see Tab.~\ref{tab:tracks-summary}).
 \1 \Colorbox{backcolour}{\lstinline|-write 2|}: all events are stored, whether they contain muons reaching the can or not. As noted above, there are other secondaries at sea level apart from muons. The types of secondary particles to be saved can be controlled with the \Colorbox{backcolour}{\lstinline|-save|} option:
   \2 \Colorbox{backcolour}{\lstinline|-save mu|} : only $\mu$ secondaries at sea level are saved.
   \2 \Colorbox{backcolour}{\lstinline|-save nu|} : only $\nu$ secondaries at sea level are saved.
   \2 \Colorbox{backcolour}{\lstinline|-save lep|} : $\mu$ and $\nu$ secondaries at sea level are saved. Here, it must be emphasised that “lep” does not indicate all leptons --- if there are any electrons or taus at sea level, they will not be stored in the output.
   \2 \Colorbox{backcolour}{\lstinline|-save all|}: all secondary particles at sea level are saved, regardless of their type.
\end{outline}

gSeaGen offers the possibility to perform format conversion of the CORSIKA output without propagation. To use gSeaGen in such a mode, one must add
\Colorbox{backcolour}{\lstinline|--corsika-only-convert true|} as an argument when executing the \emph{gSeaNuEvGen} binary. This can be combined with the \Colorbox{backcolour}{\lstinline|-write|} and \Colorbox{backcolour}{\lstinline|-save|} options, which allow reducing the amount of saved information in the same way as described above.

\subsubsection{Supported CORSIKA options}
\label{sec:supported_options}

gSeaGen is compatible with the standard format of the CORSIKA binary output files, and adding support for  simulations with (multi)thinning (THIN and MULTITHIN options; see \cite{CORSIKA-Userguide}) is foreseen. The format of the file is automatically inferred and a respective notification is printed.
The only CORSIKA compilation setting that should always be turned on is the CURVOUT option \cite{CORSIKA-Userguide}, since gSeaGen has been designed with the curved observation level in mind (see  \ref{sec:rotation}). The code will still work without it, but the results will be more incorrect, the more horizontal the shower direction.

\subsubsection{{\fontfamily{qcr}\selectfont status} values}
\label{sec:status_values}

\hfill

Several categories of particle tracks are defined and are characterised by different values of the {\fontfamily{qcr}\selectfont status} parameter, as compiled in Tab.~\ref{tab:status_values}.

\begin{table}[H]
\centering
\caption{
Summary of possible {\fontfamily{qcr}\selectfont status} values. The CORSIKA-specific ones are detailed in Tab.~\ref{tab:corsika-specific-pIDs}. Auxiliary tracks are the ones representing the mother particles and the muon creation point information (see Tab.~\ref{tab:tracks-summary}).
}
\begin{tabular}{|c||c|}
\hline 
{status} & {description} \tabularnewline
\hline 
\hline 
{-1} & {secondary not propagated or not reaching the can}\tabularnewline
\hline 
{1} & {secondary successfully reaching the can}\tabularnewline
\hline
{1001} & {initial (sea-level) state of the particle, which reached the can}\tabularnewline
\hline 
{21} & {auxiliary track (not propagated)}\tabularnewline
\hline 
{22-28} & {auxiliary track with CORSIKA-specific ID (not propagated)}\tabularnewline
\hline 
\end{tabular}{\tiny\par}

{\small{}\vspace{0.2cm}
}{\small\par}
\label{tab:status_values}
\end{table}

When processing the input from CORSIKA, neutrinos are currently assigned {\fontfamily{qcr}\selectfont status} 21, since their transport in parallel with muons is not yet supported. If a CORSIKA muon is found to have insufficient range to undergo propagation, it retains the {\fontfamily{qcr}\selectfont status} -1, but is not transported. After a muon has successfully reached the can, it is assigned a {\fontfamily{qcr}\selectfont status} 1, and its pre-propagation state from sea level is stored with a {\fontfamily{qcr}\selectfont status} of 1001.

\subsection{Weight calculation}
\label{sec:weights}

\hfill

In gSeaGen, weights are assigned to events produced with CORSIKA. Some subtle differences with respect to the weight evaluation for events produced with GENIE are related to the different origins of the CORSIKA tracks. The “intermediate weight” $w_{\mathrm{i}}$ is computed as
\begin{equation}
\left.
w_{\mathrm{i}}= A_{\mathrm{gen}} \cdot I_{\theta} \cdot I_E \cdot E^{\gamma} \cdot T_{\mathrm{gen}} \cdot \phi_{\mathrm{CR}},
\right.
\label{eq:w3_formula} 
\end{equation}

\noindent where:
\begin{itemize}
  \item $A_{\mathrm{gen}}$ [$\mathrm{m}^2$]: area of the “generation surface” (where gSeaGen starts propagation). Depending on the user choice, it may have a constant value corresponding to the area of a circle (\Colorbox{backcolour}{\lstinline|-rt can|} option) or be computed event-by-event and include more complicated shapes (\Colorbox{backcolour}{\lstinline|-rt proj|} option). The latter is recommended, as it minimises the loss of statistics due to showers missing the can.
  \item $I_{\theta}=2 \pi \cdot |\cos(\theta_{\mathrm{max}})-\cos(\theta_{\mathrm{min}})|$: angular phase-space factor, with $\theta$ being the zenith angle of the primary nucleus, i.e.\ the angle between the trajectory of the primary and the vertical direction above the detector site.  $\theta_{\mathrm{min}}$ and $\theta_{\mathrm{max}}$ correspond to the range in which CORSIKA showers are simulated (at most 0{\textdegree} and 90{\textdegree} respectively).
  \item $E \, [\mathrm{GeV}]$ is the primary cosmic ray (CR) energy.
\item $I_E \cdot E^{\gamma}= \left \{ \begin{array}{cc}
 E^{\gamma} \cdot \ln\left(\frac{E_{\mathrm{max}}}{E_{\mathrm{min}}}\right) & \mathrm{if}\,\gamma=1\\\\
E^{\gamma} \cdot \frac{E_{\mathrm{max}}^{1-\gamma}-E_{\mathrm{min}}^{1-\gamma}}{1-\gamma} & \mathrm{if}\,\gamma\neq1
\end{array} \right. $: energy phase-space factor [$\mathrm{GeV}$]. 

$E_{\mathrm{min}}$ and $E_{\mathrm{max}}$ represent the energy range of the CORSIKA simulation.
  \item $\gamma$ is the spectral index used for the energy spectrum of the CORSIKA events.
  \item $T_{\mathrm{gen}}=1\,\mathrm{s}$: this value guarantees that the production rate is in ${\mathrm{s}}^{-1}$.
  \item $\phi_{\mathrm{CR}}\,[\frac{\mathrm{m}^2}{\mathrm{s} \cdot \mathrm{sr} \cdot \mathrm{GeV}}]$: primary CR flux model. Currently, the only option in gSeaGen is the GST3 model \cite{GST}. It is possible to use an arbitrary CR primary flux by multiplying $w_{\mathrm{i}}$ by the ratio of the chosen flux to the GST3 flux.
\end{itemize}

\begin{figure}[H]
\centering
\includegraphics[scale=0.5]{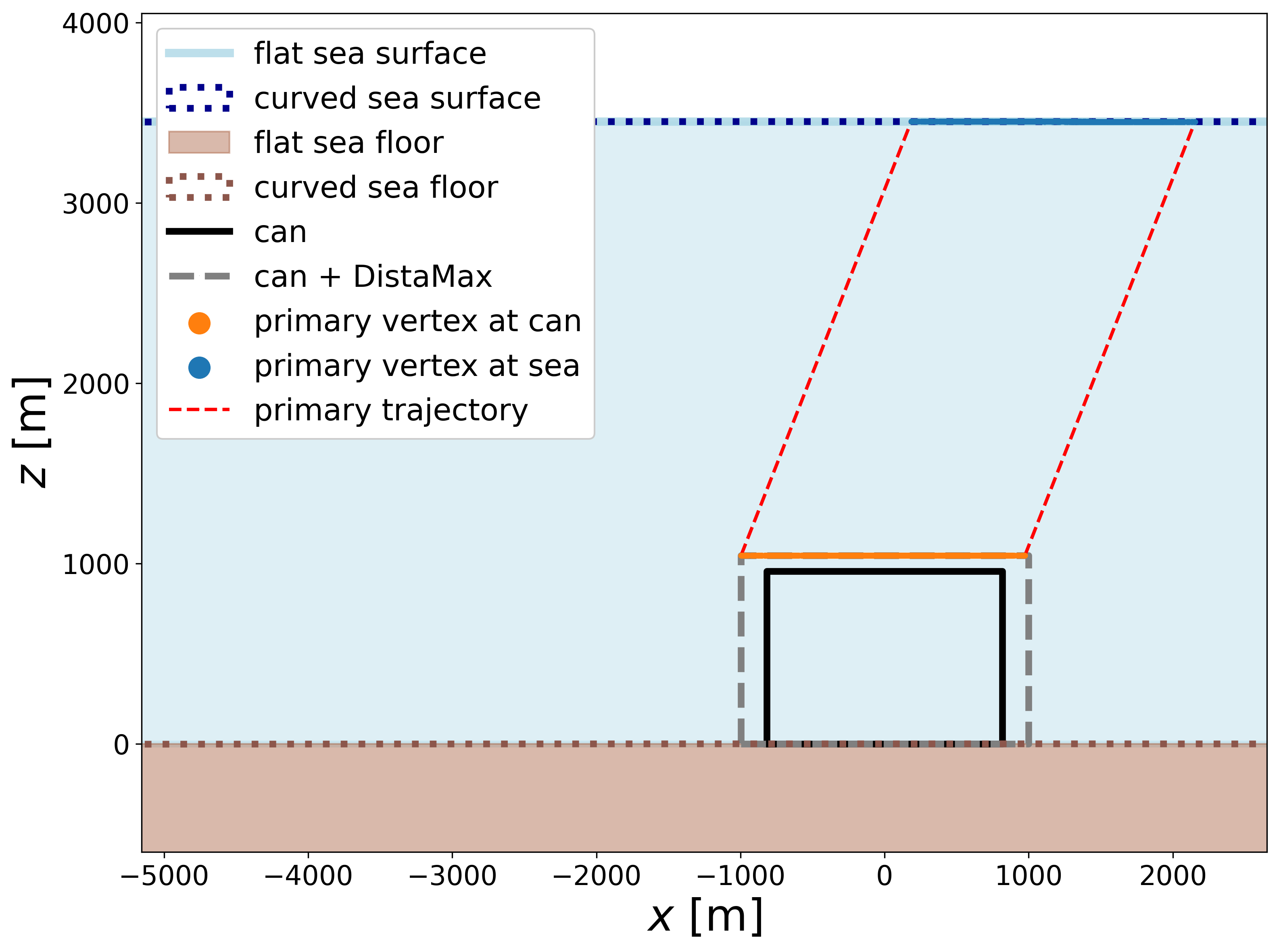}
\caption{Sketch of the projection of the can top cap area onto the sea surface (after increasing it by the appropriate \emph{DistaMax} projection onto the horizontal direction) in the side view.}

\label{fig:proj_geometry1}       
\end{figure}

\begin{figure}[H]
\centering
\includegraphics[scale=0.5]{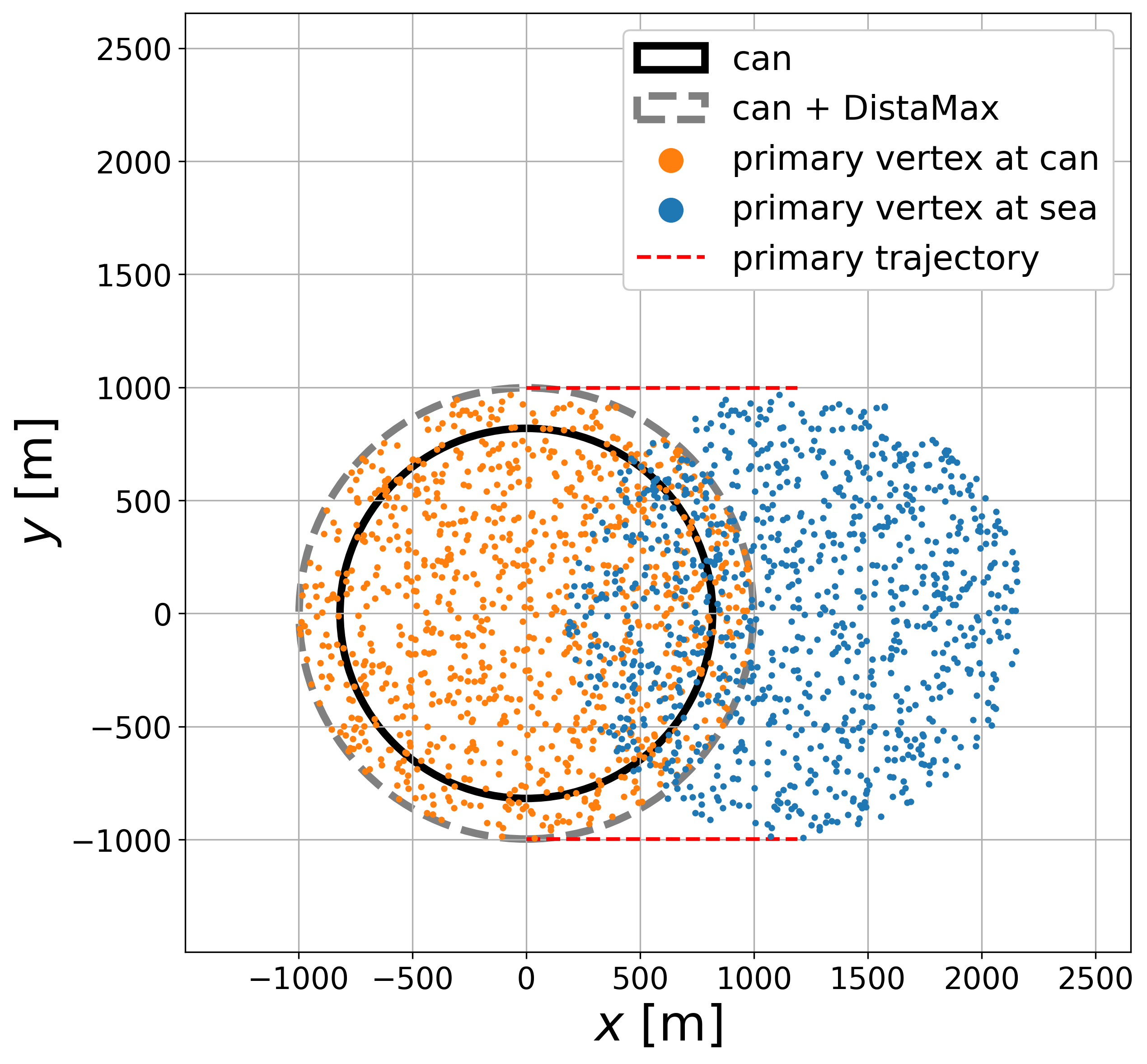}
\caption{Sketch of the projection of the can top cap area onto the sea surface (after increasing it by the appropriate \emph{DistaMax} projection onto the horizontal direction) in the top view.}
\label{fig:proj_geometry2}       
\end{figure}

\begin{figure}[H]
\centering
\includegraphics[scale=0.5]{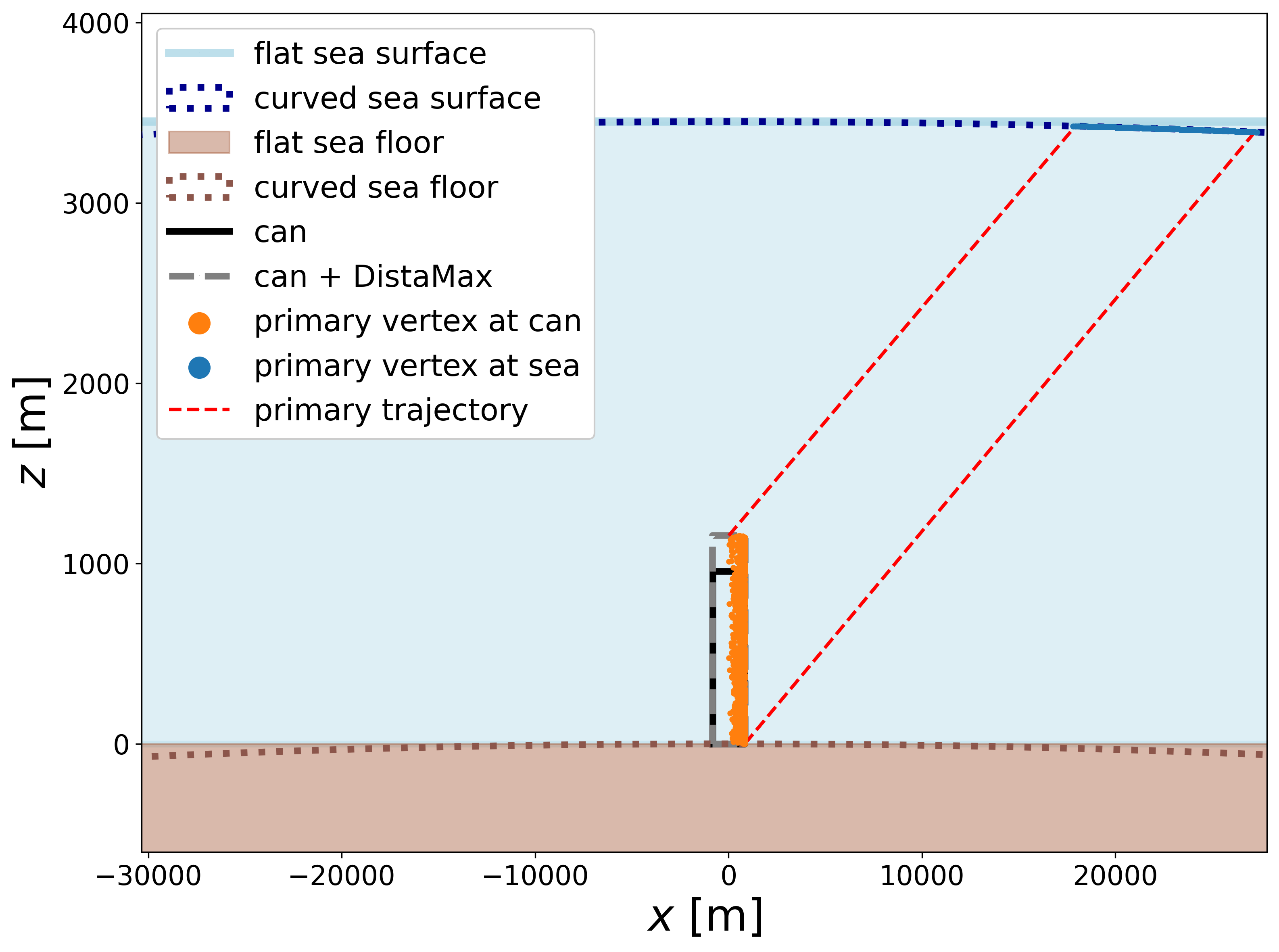}
\caption{Sketch of the projection of the can side area onto the sea surface (after increasing it by the appropriate \emph{DistaMax} projection onto the vertical direction) in the side view.}
\label{fig:proj_geometry3}       
\end{figure}

\begin{figure}[H]
\centering
\includegraphics[scale=0.5]{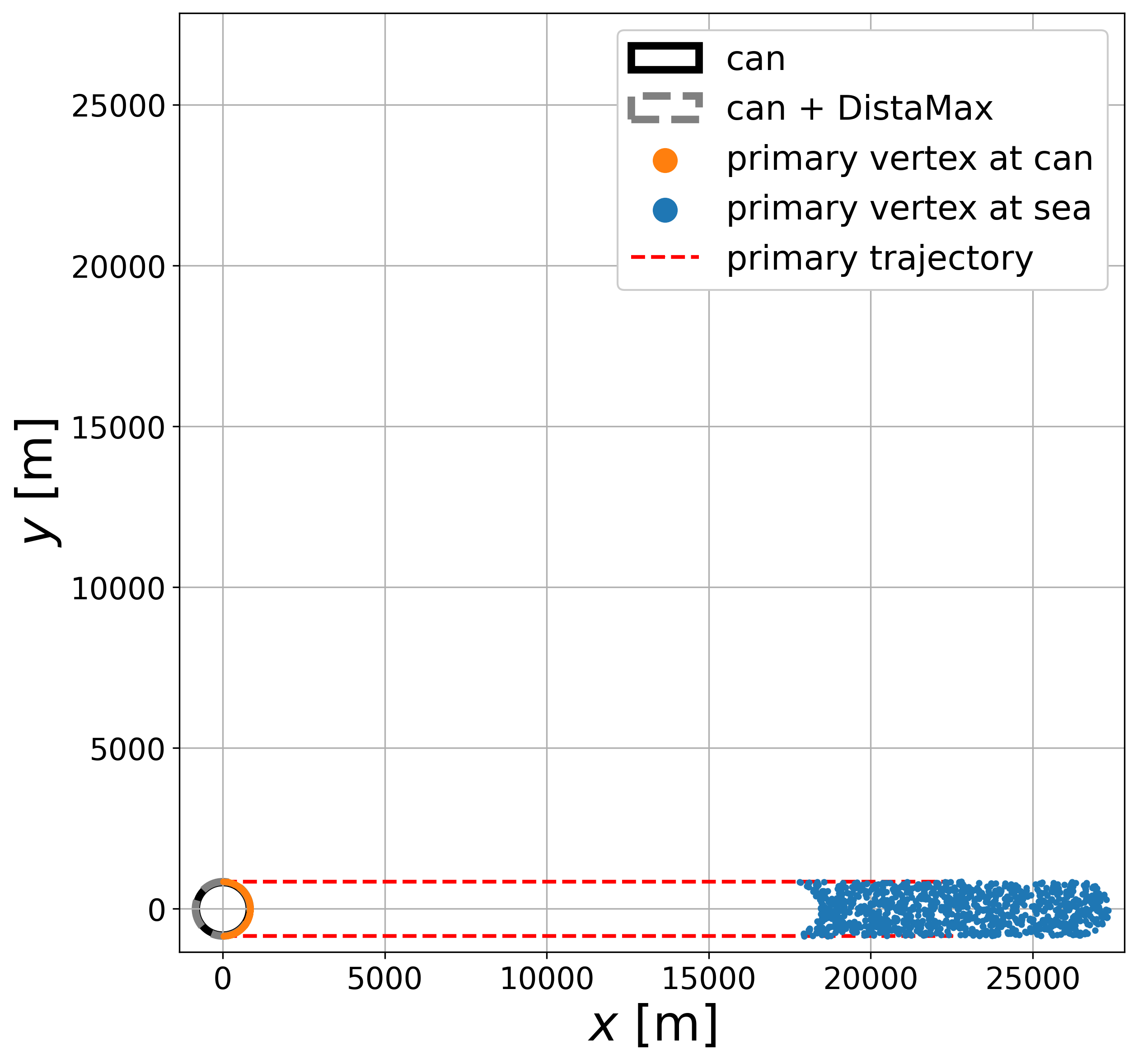}
\caption{Sketch of the projection of the can side area onto the sea surface (after increasing it by the appropriate \emph{DistaMax} projection onto the vertical direction) in the top view.}
\label{fig:proj_geometry4}       
\end{figure}

Fig.~\ref{fig:proj_geometry1} and \ref{fig:proj_geometry2} show how the area of the top cap of the can is projected onto the sea surface if the \Colorbox{backcolour}{\lstinline|-rt proj|} option is used. In Fig.~\ref{fig:proj_geometry3} and \ref{fig:proj_geometry4} the same is shown for the side area of the can. The direction of the primary CR is the main reference in CORSIKA, since all muon positions at the sea surface are defined with respect to it. In each of the four figures, the points on the can surface are sampled for a fixed primary direction, with two red dashed lines added to see the slope easily. The primary vertex in the figures refers to a position at which the extended primary trajectory intersects a certain surface,  e.g., the sea. \emph{DistaMax} (see Fig.~\ref{fig:distamax-sketch}) is the maximal lateral distance of a muon from the primary axis (computed at sea level), described in more detail in Sec.\ \ref{sec:distamax_calc}. The can radius and height are increased by the projection of \emph{DistaMax} (orthogonal to the primary direction) and so is the area $A_{\mathrm{gen}}$. This introduces an additional safety margin, which takes into account potential muon scattering. As one may note in Fig.~\ref{fig:proj_geometry3}, the curvature of the Earth plays a role in very inclined events because for a curved sea surface the distance between the sea level and the can is smaller. This effect is accounted for in gSeaGen. This includes explicit treatment of the CORSIKA CURVOUT option \cite{CORSIKA-Userguide}. Differences induced by assuming a flat or curved Earth result in a measurable effect in muon data for the KM3NeT detectors, and potentially for other neutrino telescopes as well.

The final event weight is:

\begin{equation}
\left.
w_{\mathrm{event}} = \frac{w_{\mathrm{i}} \cdot f_{\mathrm{corr}}}{n_{\mathrm{gen}}},
\right.
\label{eq:final_weight} 
\end{equation}

\noindent where $n_{\mathrm{gen}}$ is the total number of showers generated with CORSIKA\footnote{It is important that each primary is counted separately.}
and $f_{\mathrm{corr}}$ is the correction factor needed to account for additional chances given to the event to reach the can (see Sec.\ \ref{sec:retrying_showers}), with a default value of 1 (no retrying). The weight correction factor $f_{\mathrm{corr}}$ can be evaluated using the gWeightCorsika application, provided along with gSeaGen and located in:
\begin{lstlisting}[language=bash]
/src/tools/gWeightCorsika/
\end{lstlisting}
The program evaluates $f_{\mathrm{corr}}$ as a function of energy and direction of the primary particle by comparing unweighted interpolated histograms of non-retried events with histograms containing all events. The predicted number of events per second is obtained by weighting the simulation with $w_{\mathrm{event}}$.

\subsubsection{Evaluation of \emph{DistaMax}}
\label{sec:distamax_calc}

\hfill

\begin{figure}[H]
\centering
\includegraphics[scale=0.4]{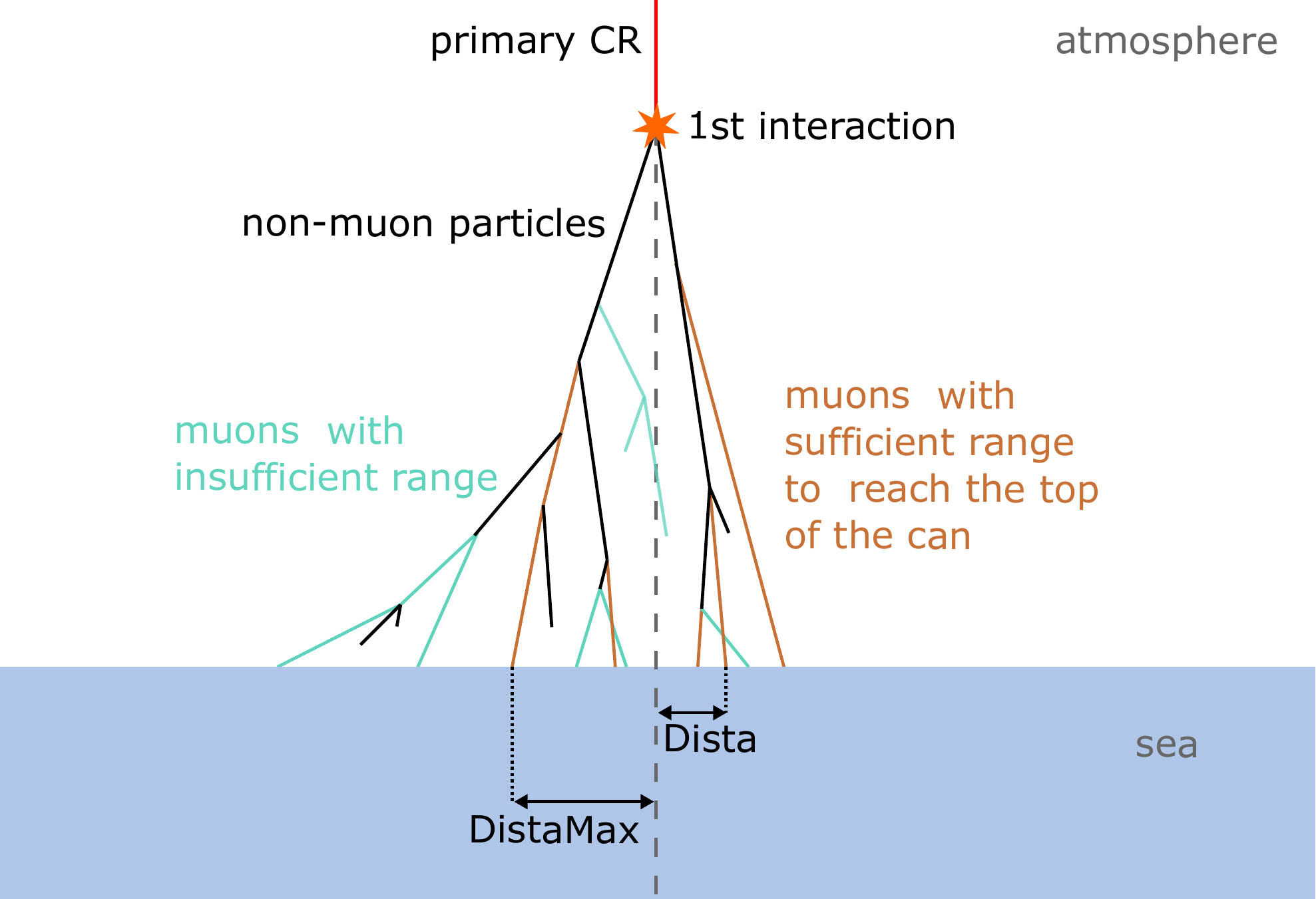}
\caption{Sketch illustrating the meaning of \emph{DistaMax} using the example of a vertical shower. The value of \emph{DistaMax} is always evaluated for muon positions at the sea surface.}
\label{fig:distamax-sketch}       
\end{figure}

\begin{figure}[H]
\centering
\includegraphics[scale=0.5]{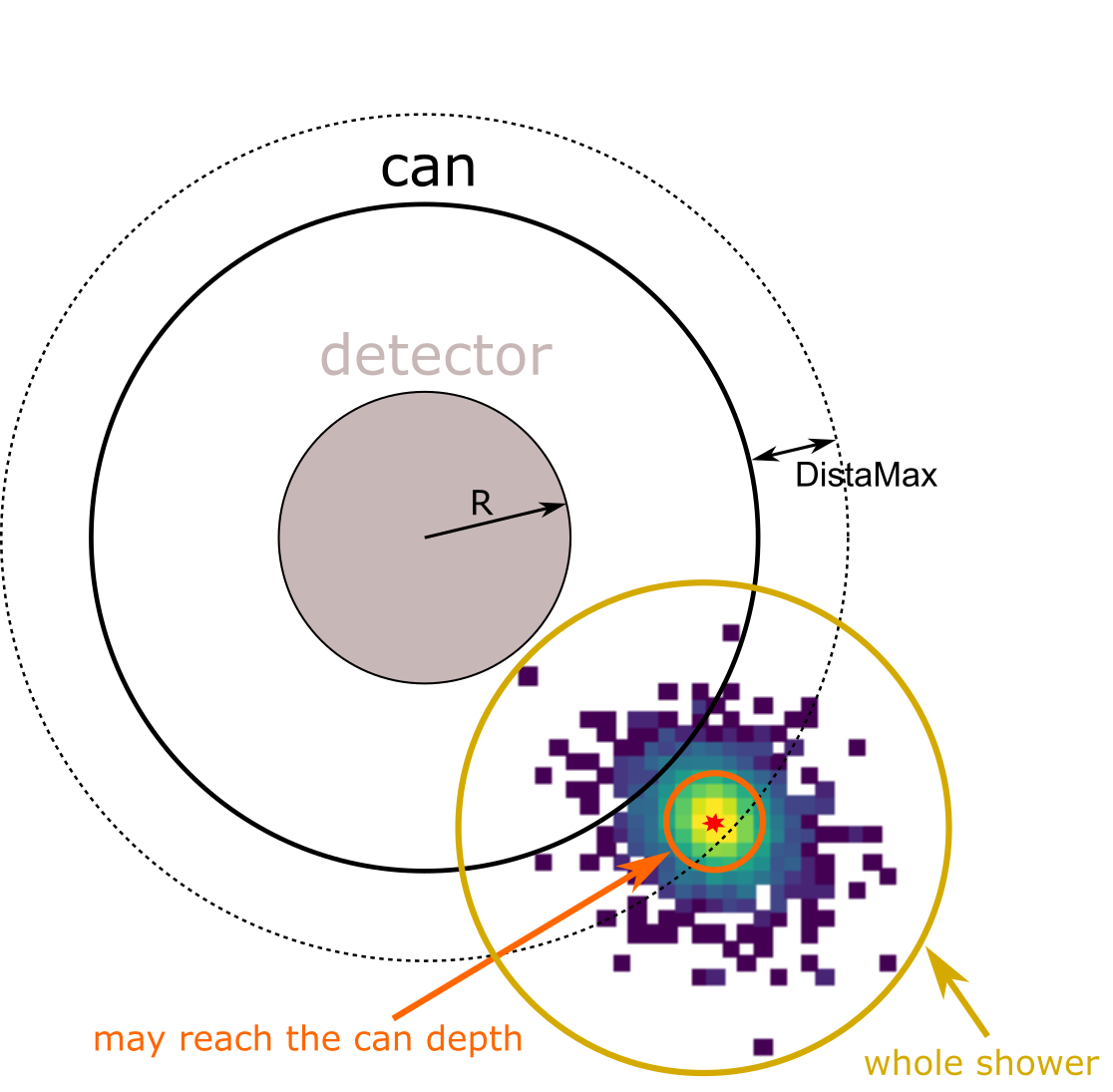}
\caption{Top view of the geometry of shooting the shower at the can. The drawing assumes a perfectly vertical shower for simplicity; however, an actual histogram of an EeV proton shower simulated with CORSIKA (without thinning) from a KM3NeT CORSIKA simulation has been used, showing the muon density at the sea level. The yellow and orange circles are sketched to illustrate the qualitative effect of muon selection by range, which is effectively a cut on muon energy. The more energetic muons are typically closer to the shower axis since they are harder to deflect. This leads to the narrowing of the shower when the muons with insufficient range are discarded.}
\label{fig:geometry-shooting}       
\end{figure}

The value of \emph{DistaMax} is a measure of the lateral spread of muons around the primary axis, as illustrated in Fig.~\ref{fig:distamax-sketch}. Its value is used to additionally increase the can size and take into account the cases where a shower intercepts the can, but only with a few laterally scattered muons (or even a single muon). The computation of \emph{DistaMax} has been optimised, and it is evaluated considering only muons with a sufficient range (which depends on the muon energy) to reach the can (see Fig.~\ref{fig:geometry-shooting}). As a result, up to 4 times more showers from primaries with EeV energies can be processed (64\% instead of 15\% if all muons are considered). This is a particularly important feature when using CORSIKA simulations, since high-energy showers are extremely costly in terms of storage and computational time.

To account for the rare cases where a muon trajectory would not intersect the can, but the muon scatters under a large angle and ends up entering the can, the \emph{DistaMax} value itself is enlarged by increasing the lateral distance of each muon by an individually computed value. This lateral deflection safety margin is based on a two-step fit of muon deflection as a function of slant depth and energy, produced using the PROPOSAL propagator \cite{PROPOSAL} (see \ref{sec:muon_range_deflection}).

\subsection{Additional functionalities}

\hfill

Several functionalities have been introduced to improve the CORSIKA simulation statistics and facilitate the processing of CORSIKA files.

\subsubsection{Repetition of shower propagation}
\label{sec:retrying_showers}

\hfill

The possibility of giving muon bundles additional chances to reach the can is available using the \Colorbox{backcolour}{\lstinline|-chances|} option. For example, to allow the propagation of each shower to be retried up to 100 times (default is 0), the argument \Colorbox{backcolour}{\lstinline|-chances 100|} is used. If a bundle arrives at the can, no further attempts are made. The implementation of propagation with retrying focusses on recovering statistics from failed events. However, this comes at the price of a non-uniform distortion of event distributions. To account for it, a correction factor $f_{\mathrm{corr}}$ must be applied, as indicated in Equation \ref{eq:final_weight}.

%
%

\subsubsection{Reduction of verbosity}
\label{sec:reduced_verbosity}

\hfill

Low-energy CORSIKA simulations may require generating millions of showers per run since many showers will simply not reach the can. Such a large number of showers could cause the gSeaGen log files to become huge. This issue is addressed with the \Colorbox{backcolour}{\lstinline|--corsika-less-verbose 1|} option, which prints information every 100,000th (instead of 500th) generated event and will completely skip information about reading the run and event header.

\subsubsection{Muon range tolerance}
\label{sec:muon-range-tol}

\hfill

To increase the low-energy simulation statistics, an option is available to allow for a more gracious treatment of low-energy muons. For example, by adding \Colorbox{backcolour}{\lstinline|--muon-range-tolerance 50.0|} when executing the code, the tolerance margin for the muon range is increased by $50\,$m, allowing the muons with a small probability of reaching the can to get there (muons with an insufficient range are not propagated at all).

\section{Propagation}
\label{sec:propagation}

This section offers a qualitative description of the muon propagation routine of gSeaGen. 

\subsection{Propagation procedure}

The propagation is performed following the steps listed below:\

\begin{enumerate}
    \item {Reading the information from the muon tracks and initial orientation of the shower:} 
    
    \hfill
    
    \begin{enumerate}
        \item Each CORSIKA shower is read and \emph{DistaMax} is evaluated, taking into account the safety margin for muon scattering (see Sec.\ \ref{sec:weights}).
            
        \hfill
    
        \item The starting point of the shower trajectory is located above the centre of the can, at a vertical height corresponding to the observation level (primary vertex). The muon positions are defined relative to the primary vertex and lie on the observation level. Each shower has to be reoriented such that its trajectory will intersect the enlarged can (see \ref{sec:rotation}). There are two possibilities:
            
        \hfill
    
        \begin{itemize}

            \item \Colorbox{backcolour}{\lstinline|-rt proj|}: sampling showers on the can surface area projected onto the curved observation level surface. This is the recommended option for the propagation of CORSIKA muons. The projection works as follows:

            \hfill
            
            \begin{itemize}
                \item Projected areas of the top cap ($A_{\mathrm{top}}$) and side of the can ($A_{\mathrm{side}}$) are computed.

                \hfill
                
                \item Whether the shower axis will aim at the top cap or on the side of the can is decided randomly, with the probabilities proportional to $A_{\mathrm{top}}$ and $A_{\mathrm{side}}$ respectively.

                \hfill
                
                \item Depending on the result of the selection, a random point is picked on the surface of the top cap of the can or on the side of the can.

                \hfill
                
                \item This point is traced back to the observation level using the direction of the primary.

                \hfill
                
                \item The entire shower (including tracks which do not undergo propagation) is moved such that the primary vertex is located at that point (for details, see \ref{subsec:rotation-around-Earth}).

                \hfill
                
            \end{itemize}

            \hfill

            \item \Colorbox{backcolour}{\lstinline|-rt 'can'|}: the starting position of the shower is randomly selected on a disk covering the projection of the detector can onto a plane perpendicular to the primary direction. A detailed description of this option is given in \cite{gSeaGen-2020}. This is not a recommended strategy for CORSIKA events due to its inefficiency. The directions of showers sampled on a part of the area of the disk will not intersect the can, because the projected area of the can (a cylinder) is not a disk (unless the shower is perfectly vertical).
            
        \end{itemize}
    \end{enumerate}
    \item {First step of propagation:}

    \hfill
    
    \begin{enumerate}
        \item The muons whose ranges are sufficiently long (see Sec.\ \ref{sec:muon_range_deflection}) are propagated from the observation level to $h_{\mathrm{can}}$ (height of the top cap of the can).

        \hfill
        
        \item For each muon, a check is made to see whether it has already hit, missed, or has not reached the can yet.

        \hfill
        
        \begin{itemize}
            \item If there are muons in the can or pointing at it but not inside it yet (they point at the side of the can), propagation proceeds to the next stage.

            \hfill
            
            \item In the case when no muon reaches the can and none has a chance to do so in the second stage, but some muons just missed the can, the shower is either randomly rotated (\Colorbox{backcolour}{\lstinline|-rt proj|} option) or horizontally shifted (\Colorbox{backcolour}{\lstinline|-rt 'can'|} option) into the can. This is counted as retrying and is only done if \Colorbox{backcolour}{\lstinline|-chances|} is used (see Sec.\ \ref{sec:retrying_showers}).

            \hfill
            
            \item If there is no point in further propagation or shifting, the shower can be retried by starting over. The event is reset and shifted to a new position at the observation level. This will only be done if the \Colorbox{backcolour}{\lstinline|-chances|} option is used.

            \hfill
            
            \item Should everything fail, the event remains unsuccessful, and the code moves to the next one.

            \hfill
            
            \item For successful events with nothing left to propagate (i.e., all muons either hit the top of the can or failed to hit the can), the weights are computed, and the event is written into the output file.

            \hfill
            
        \end{itemize}
    \end{enumerate}
    \item {Second step of propagation from the level corresponding to the top of the can, $h_{\mathrm{can}}$, to an interception point on the can. The propagation of muons is resumed at the “top of the can” level and there are three possible cases:}

    \hfill
    
    \begin{enumerate}
        \item No muon reaches the can, but the specified allowed number of retries (see Sec.\ \ref{sec:retrying_showers}) has not been exceeded yet. In such a case, the event may be reconsidered for a further propagation attempt.

        \hfill
        
        \item No muon reaches the can and there are no attempts left --- the code moves to the next event.

        \hfill
        
        \item There is at least one muon in the can, meaning that the propagation was successful. Subsequently, the weight of the event is computed and the propagated event is saved in the output file. For options to control the amount of saved information, see Sec.\ \ref{sec:track-info}.

        \hfill
        
    \end{enumerate}
\end{enumerate}

\subsection{Customization of muon transport}

Although the defaults of gSeaGen are optimised for the specific case of KM3NeT, the simulation may be adjusted in several ways. The medium in which the propagation is performed does not have to be water and does not have to start at the sea level (but does have to match the observation level from the CORSIKA files used). If PROPOSAL is used for particle transport, it is also possible to have multiple media of various shapes and with variable densities \cite{PROPOSAL}. In the case of the basic built-in routine --- PropaMuon, this is not possible.

The propagation steps can also be controlled using the following two options:
\begin{itemize}
    \item \Colorbox{backcolour}{\lstinline|--approach-dist|} --- sets the distance from the can to which the first propagation step will be done (in meters),
    \item \Colorbox{backcolour}{\lstinline|--propagation-step|} --- sets the propagation step size (in meters).
\end{itemize}
The reason for the introduction of these is that if propagation is done in a single step, it might happen that the muon will scatter laterally and enter too far into the can. This is undesirable because gSeaGen does not simulate the light emission, and hence having muons too far inside the can will cause underestimation in the light simulation at the subsequent simulation stage.

\section{Summary}

Several features have been added to the gSeaGen code to enable the processing of muons from air showers simulated with CORSIKA. For this purpose, direct readout of CORSIKA output has been implemented, as well as the possibility to either directly convert it to a different data format or to perform propagation of simulated atmospheric muons to a specified detector. The user of gSeaGen can control how propagation is carried out and what information is deemed worth saving. The muon propagation routine includes many optimisations, which boost both the accuracy of the muon transport and the speed of the code. Options facilitating the reuse of generated showers have been added. These developments were made with simulations of the KM3NeT neutrino telescopes primarily in mind, but other underground, -water, or -ice experiments may profit from them as well.

The efforts towards a more efficient and complete treatment of CORSIKA air shower simulations continue. The improvements foreseen in the future include, but are not limited to:
\begin{itemize}
    \item propagation of neutrinos together with muons,
    \item upgrade to a newer version of PROPOSAL \cite{PROPOSAL},
    \item re-evaluation and further improvement of the muon lateral scattering fit, including a possibility to use a custom fit (prepared for the specific media used in the simulation),
    \item adding the KM3NeT data format as one of the possible output format choices,
    \item treatment of the THIN and MULTITHIN CORSIKA options, in particular regarding their impact on the DistaMax parameter,
    \item investigation of the possibility to directly utilize public libraries of air showers simulated with CORSIKA,
    \item inclusion of the possibility to simulate non-standard configurations of incident cosmic rays, such as cosmic ray ensembles \cite{CREDO},
    \item adding jupyter notebooks with examples.
\end{itemize}
User feedback is welcome and should preferably be expressed via issues in the official \href{https://git.km3net.de/opensource/gseagen}{GitLab repository}.

\appendix
\section{}
\label{sec:appendix}

\subsection{Muon range and deflection in water}

\hfill

\label{sec:muon_range_deflection}

\begin{figure}[H]
\centering
\includegraphics[scale=0.5]{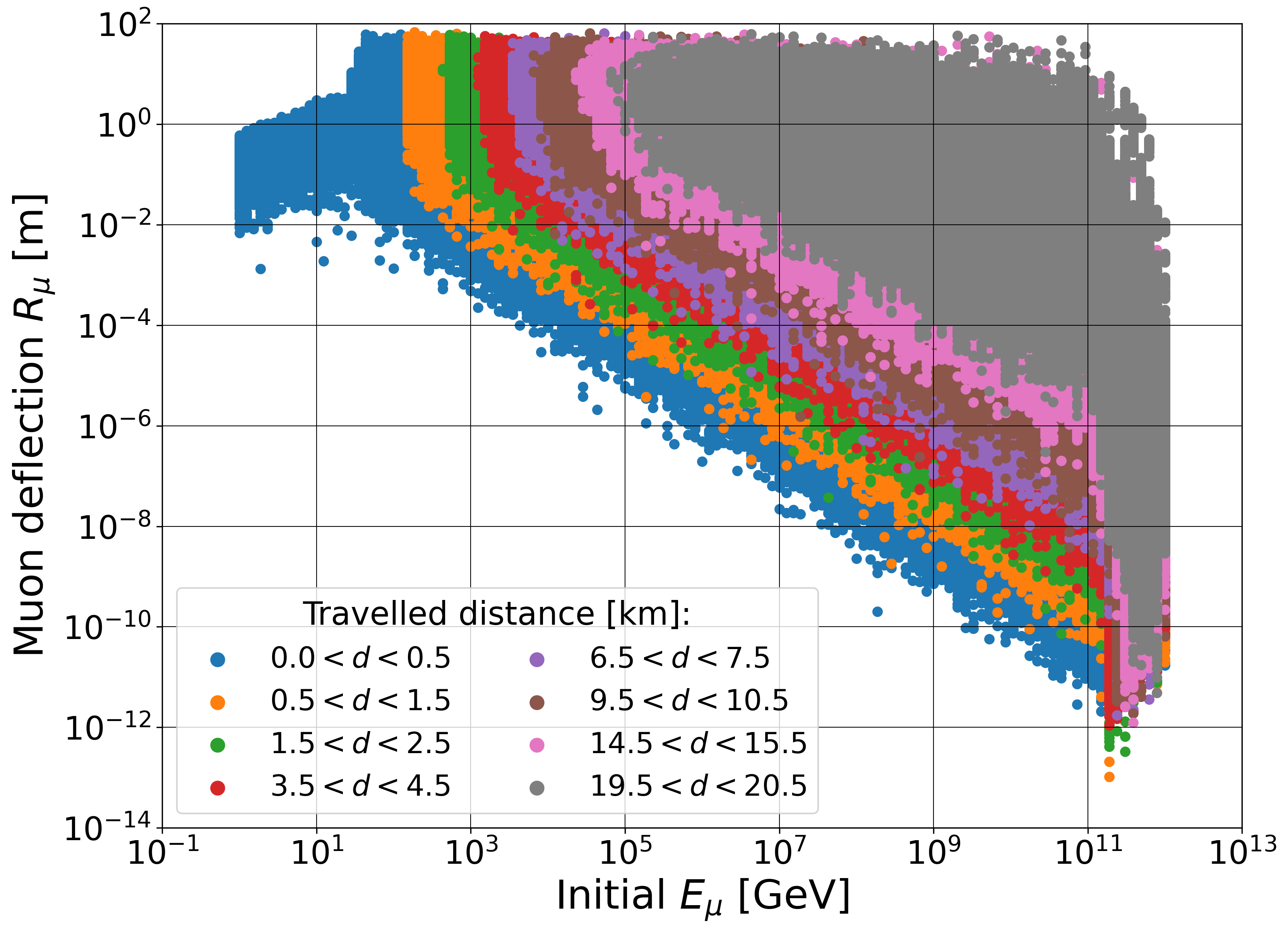}
\caption{Lateral deflection of muons $R_{\mu}$ as a function of their initial energy $E_{\mu}$ and distance $d$ travelled in seawater.}
\label{fig:deflection}       
\end{figure}

During investigation of possible improvements of the gSeaGen code, the lateral deflection of muons when travelling through water was studied. One of the results of the study is the estimation of the lateral muon deflection as a function of its initial energy and distance travelled (see Fig.~\ref{fig:deflection}), obtained by using a modified gSeaGen code with PROPOSAL (v6.1.5) as muon propagator. For this study, no CORSIKA input was used. Instead, all muons were injected with a vertically downward momentum and followed through seawater until they stopped. This scenario simplified the evaluation of the deflection to $R_{\mu}=\sqrt{\Delta x^{2}+\Delta y^{2}}$, where $\Delta x$ and $\Delta y$ are the muon lateral displacements with respect to the initial muon position. The initial muon energies were sampled in the range between 1$\,$GeV and 1$\,$ZeV ($10^{12}\,$GeV), with 1000 muons generated for each energy and the muon deflection recorded every 100$\,$m. For the highest energies, the longest distances travelled by the muons exceeded 100$\,$km. For comparison, the geometrical limit for the travelled distance in CORSIKA simulation for KM3NeT is $d_{\mathrm{max}}=60.46\,\mathrm{km}$, which can be computed from 
\begin{equation}
    {d_{\mathrm{max}}=R_{\mathrm{Earth}} \cos (\theta_{\mathrm{max}})+\sqrt{R_{\mathrm{Earth}}^2 \cos^2 (\theta_{\mathrm{max}})-D_{\mathrm{max}}^2-2R_{\mathrm{Earth}}D_{\mathrm{max}}}},
\label{eq:dmax}
\end{equation}
where $R_{\mathrm{Earth}}=6371.315\,$km is the average radius of the Earth (the value used in CORSIKA), $D_{\mathrm{max}}=3500\,$m is the maximum depth (depth of the ARCA site) \cite{KM3NeT-LoI-2.0} and $\theta_{\mathrm{max}}=87\text{°}$ is the maximum zenith angle considered in CORSIKA simulation for KM3NeT. Larger zenith angles are not simulated, since the muon flux due to extremely horizontal showers is negligible and would require an enormous number of generated events to obtain some particles at the can. Notably, this will no longer apply in the case of neutrinos.

Various physical effects can be observed in Fig.~\ref{fig:deflection}. The vertical left edges for different travelled distances demonstrate that there is a threshold energy required for a muon to be able to travel a particular distance. The upper limit on the lateral deflection appears to be roughly 100$\,$m and shows a weak dependence on the initial muon energy above 100$\,$GeV. In the region of $1-50\,$GeV, the muon ranges are relatively short, that is, the muon experiences less scattering before it stops, which limits the maximal achievable lateral deflection. However, the large minimal value demonstrates that the scattering at low energies significantly affects the muon trajectory. Moving towards the initial $E_{\mu}$ of about 100$\,$GeV, the muon range stops to play a role and the deflection reaches the global maximum. Upon further increase of the energy, tiny lateral deflections become more probable because muons have stronger Lorentz boosts and since high-momentum particles are harder to deflect. This is reflected in the downward slope in Fig.~\ref{fig:deflection}.

\begin{figure}[H]
\centering
\includegraphics[scale=0.45]{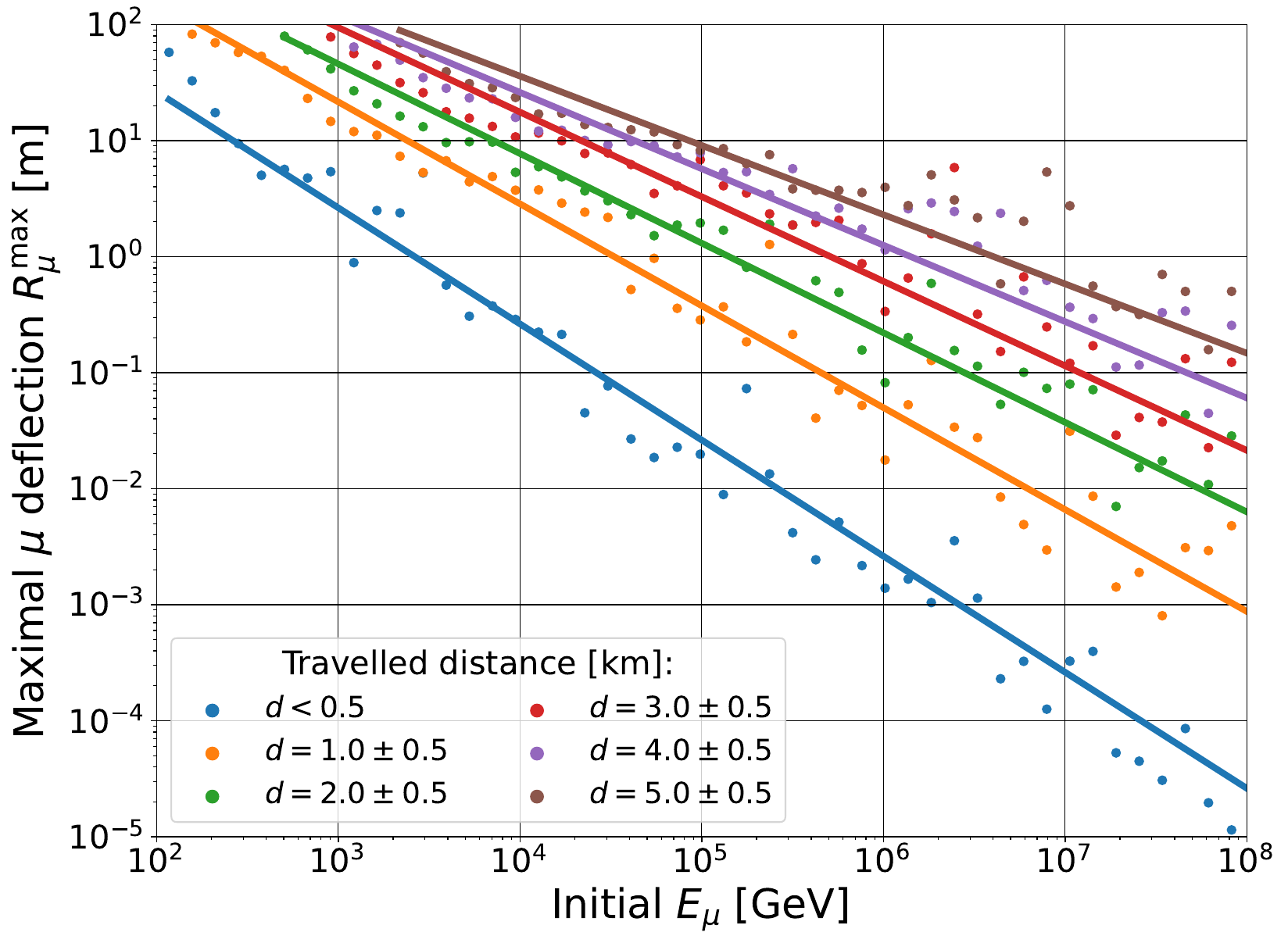}
\caption{Maximal lateral muon deflections fitted as a function of $E_{\mu}$. The fitting function is $R(E_{\mu})=10^{a_d\cdot\log_{10}(E_{\mu})+b_d}$.}
\label{fig:fit_examples}       
\end{figure}

\begin{figure}[H]
    \centering
    \includegraphics[scale=0.45]{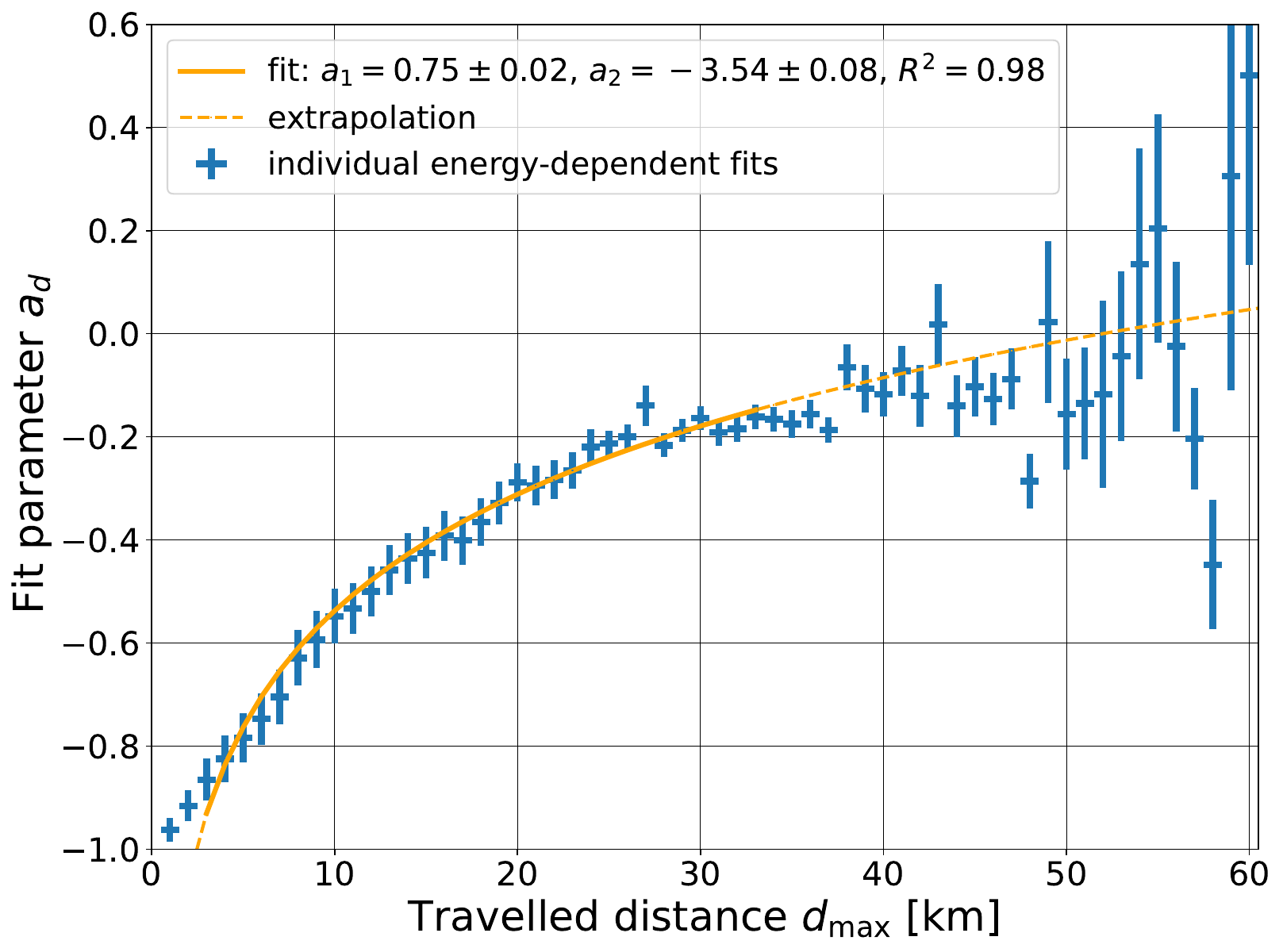}
    \caption{Dependence of the energy-dependent fit parameter of the maximal lateral muon deflection $a_d$ on the distance travelled. The fitting function is $a_d = a_1 \cdot \log_{10}(\frac{d_{\mathrm{max}}}{\mathrm{m}})+a_2$ and $R^2$ is the coefficient of determination.}
    \label{fig:fitting_the_fit_1}
\end{figure}

\begin{figure}[H]
    \centering
    \includegraphics[scale=0.45]{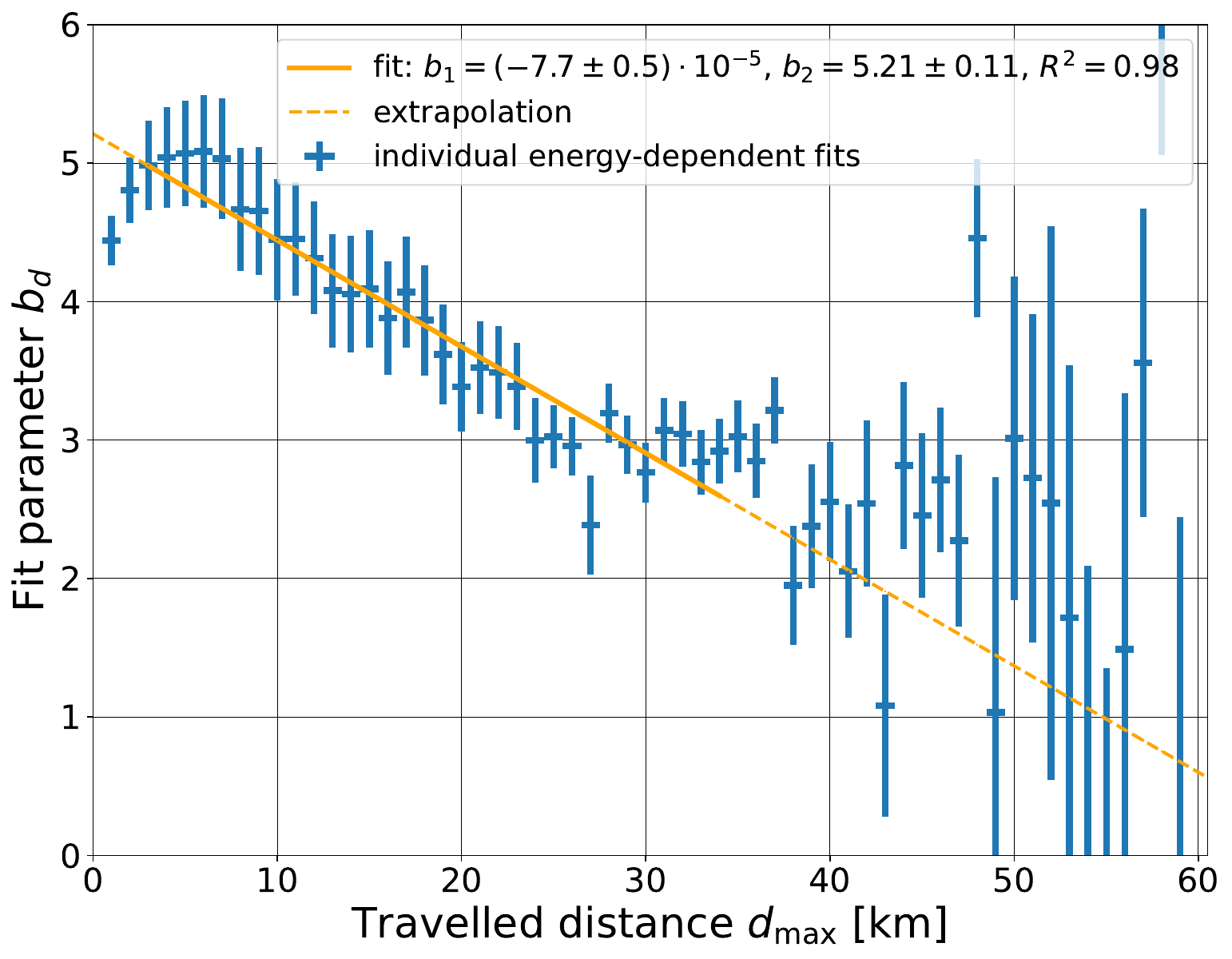}
    \caption{Dependence of the energy-dependent fit parameter of the maximal lateral muon deflection $b_d$ on the distance travelled. The fitting function is $b_d = b_1 \cdot \frac{d_{\mathrm{max}}}{\mathrm{m}}+b_2$ and $R^2$ is the coefficient of determination.}
    \label{fig:fitting_the_fit_2}
\end{figure}

It is possible to fit the maximal deflections (here approximated with the average deflection increased by 10 standard deviations: $R_{\mathrm{avg}}+10{\cdot} R_{\mathrm{std}}$) linearly in log$-$log scale as a function of $E_{\mu}$. Examples of such fits are shown in Fig.~\ref{fig:fit_examples}. Moreover, also the distance dependence has been successfully parameterised, as it turned out that the linear fit parameters from the $E_{\mu}$-dependent fit ($a_d$ and $b_d$, as defined in the caption of Fig. \ref{fig:fit_examples}) vary linearly (in appropriate scales) with the distance $d$, as shown in Fig.~\ref{fig:fitting_the_fit_1} and \ref{fig:fitting_the_fit_2}. All this allowed for a fast computation of the maximal expected deflection for each muon:

\begin{equation}
    R_{\mathrm{max}}=10^{\left [ a_1 \cdot \log_{10}\left(\frac{d_{\mathrm{max}}}{\mathrm{m}}\right)+a_2 \right]\cdot\log_{10}(\frac{E_{\mu}}{\mathrm{GeV}})+b_1 \cdot\frac{d_{\mathrm{max}}}{\mathrm{m}}+b_2}\,[\mathrm{m}],
\label{eq:parametrisation}
\end{equation}

\noindent where 
$d_{\mathrm{max}}=\frac{\mathrm{site\,depth}}{\cos\theta}$, 
$a_1=0.75\pm0.02$, 
$a_2=-3.54\pm0.08$,  
$b_1=(-7.7\pm0.5)\cdot 10^{-5}$, and 
$b_2=5.21\pm0.11$. Eq.~\ref{eq:parametrisation} has been implemented in the gSeaGen muon propagation routine (Sec. \ref{sec:propagation}). In the future releases it is planned to make the fit of the maximal muon deflection customisable, i.e. let the users provide their own, depending on the medium in which they want to propagate the muons. This would come with a step-by-step code example, illustrating how to produce it.

\label{sec:mu_range_tolerance}

\subsection{Reorientation of CORSIKA showers}

\hfill

\label{sec:rotation}

The gSeaGen code ensures that an event will hit the detector by moving it around, causing its trajectory (inferred from the direction before propagation) to intersect the can. A vital part of this work was a fundamental rework of the CORSIKA shower reorientation in gSeaGen. Previously, the showers were only horizontally shifted (in the {$x$-$y$} plane) assuming a flat geometry. Here, a more accurate approach was introduced taking into account the curvature of the Earth. This plays a significant role for horizontal showers, where the differences between the flat and curved scenarios could reach hundreds of metres. 

The geometry of the problem was assumed to be symmetric around the centre of the Earth (see Fig.~\ref{fig:sketch_curved_atmosphere_rotation}). The shape of the Earth was approximated as a perfect sphere with radius $R_{\mathrm{Earth}}=6371.315\,$km, matching the CORSIKA simulation \cite{CORSIKA-Userguide}. This allowed rotating showers around the Earth without changing the altitude. The only requirement for the rotation was that the shower trajectory, evaluated on the basis of the initial position and direction of the primary, should intersect the can. When processing a CORSIKA output file, the initial shower position is always at sea level at ${\color{violet}\bullet} \, \left(0,0,{\color{blue}D}\right)$ (\textcolor{violet}{violet point} in Fig.~\ref{fig:sketch_curved_atmosphere_rotation}), where ${\color{blue}D}$ is the depth at which the detector is located. The origin of the coordinate system is placed at the centre of the base of the can. The shower in its original orientation will not intersect the can unless it happens to be vertical, hence it has to be rotated. In addition, including such a random rotation is necessary to avoid a directional bias in the simulation.
\subsubsection{Computation of the required rotation}
\label{subsec:Computing-the-needed-rotation}

\hfill

\begin{figure}[H]
\centering
\centering
\includegraphics[width=\textwidth]{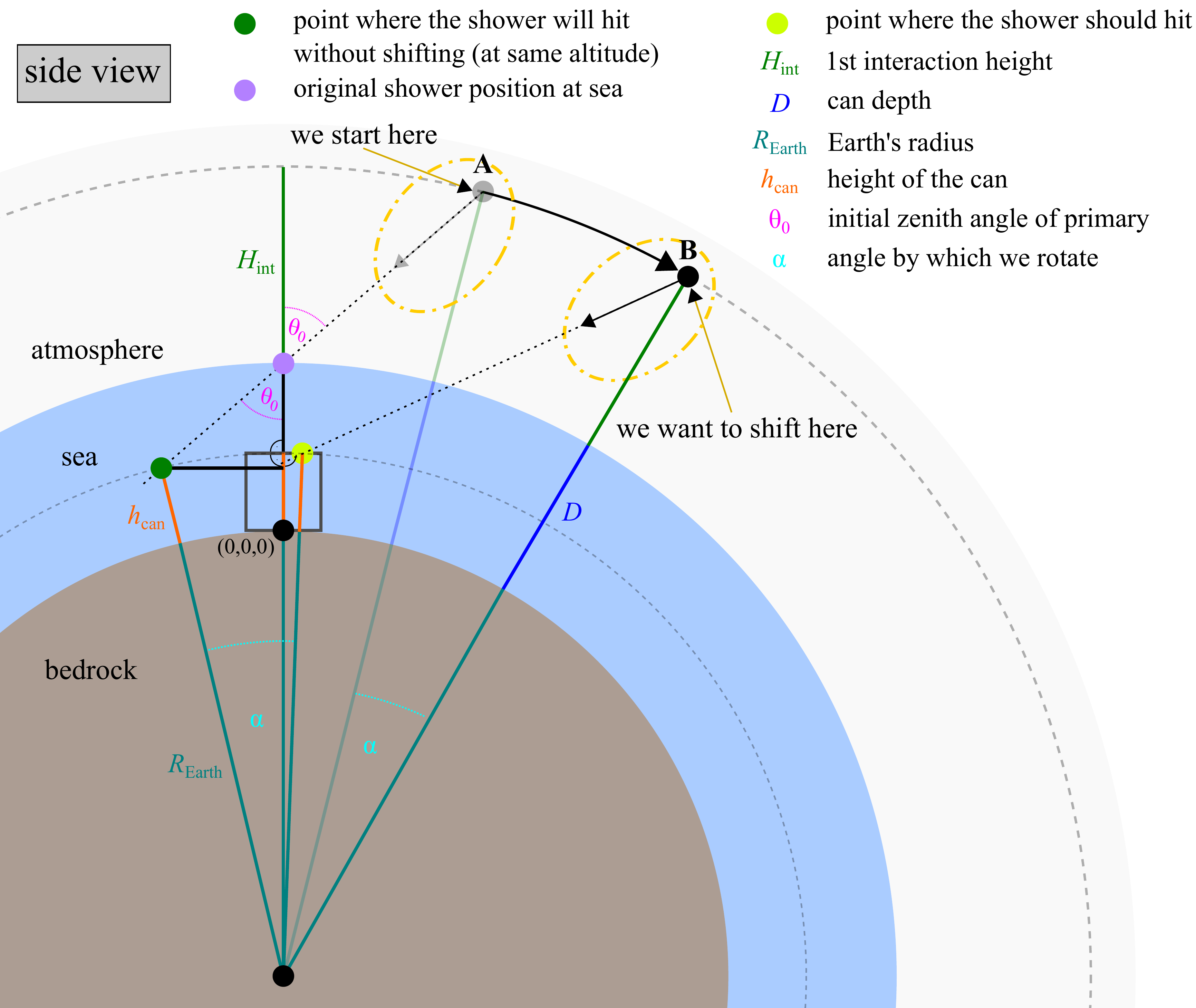}
\caption{Sketch of the geometry of shower reorientation, shown from the side.}
\label{fig:sketch_curved_atmosphere_rotation}       
\end{figure}

\begin{figure}[H]
    \centering
    \includegraphics[width=0.5\textwidth]{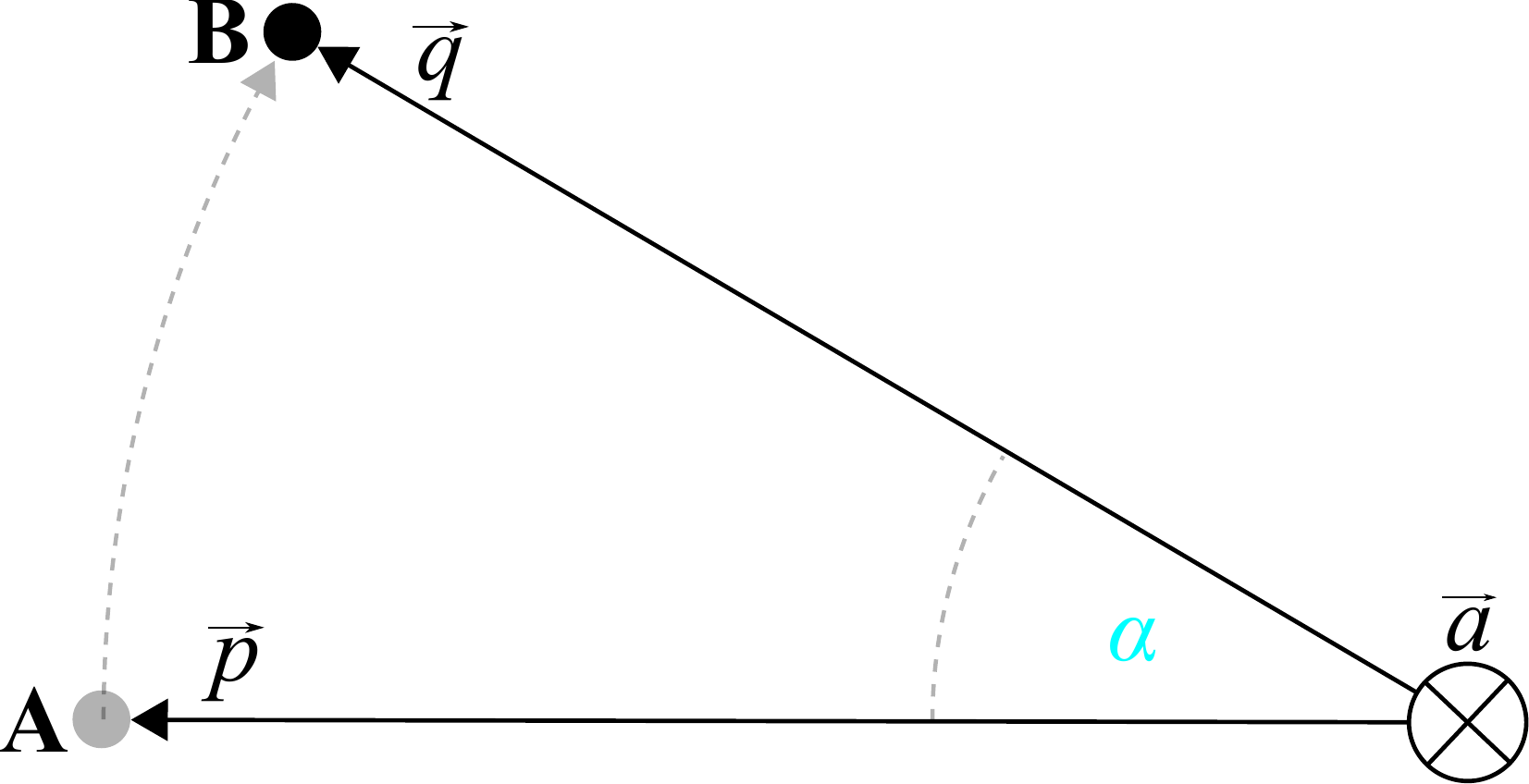}
    \caption{Sketch of the vector $\vec{p}$ rotated onto the vector $\vec{q}$ by an angle ${\color{cyan}\alpha}$ around the axis of rotation defined by the orthogonal vector $\vec{a}$.}
    \label{fig:axis-of-rotation}
\end{figure}

To evaluate the rotation matrix $M$, required for a shower trajectory to intersect a given point $\textcolor{electriclime}{\bullet} \left(x_{\mathrm{can}},y_{\mathrm{can}},z_{\mathrm{can}}\right)$, one must follow the steps:

\begin{enumerate}
    \item {
        The point $\textcolor{darkgreen}{\ensuremath{\bullet}} \left(x_{\mathrm{p}},y_{\mathrm{p}},z_{\mathrm{p}}\right)$, where the initial primary trajectory intersects with the sphere of radius $r={\color{teal}R_{\mathrm{Earth}}}+z_{\mathrm{can}}$, approximating the Earth (see Fig.~\ref{fig:sketch_curved_atmosphere_rotation}) must be found. It is sufficient to find the distance to that point since the direction $\left(c_{x,}^{\mathrm{p}},c_{y}^{\mathrm{p}},c_{z}^{\mathrm{p}}\right)=\left(c_{x,}^{\mathrm{sea}},c_{y}^{\mathrm{sea}},c_{z}^{\mathrm{sea}}\right)$  is already known. Here, a shorthand notation for directional cosines is used: $\cos{\theta_i} = c_i$,  where $i \in  (x,y,z)$. The sought distance $d_{\mathrm{intersect}}$ is the solution to the quadratic equation
        \begin{equation}
            {a\cdot d_{\mathrm{intersect}}^{2}+b\cdot d_{\mathrm{intersect}}+c=0},
        \end{equation}
        with the coefficients: 
        
        \begin{equation}
            a=1,    
        \label{eq:a}
        \end{equation}
        
        \begin{equation}
            b=2\cdot\left(c_{x}^{\mathrm{p}}x_{1}+c_{y}^{\mathrm{p}}y_{1}+c_{z}^{\mathrm{p}}z_{1}\right),
        \label{eq:b}
        \end{equation}
        
        \begin{equation}
            c=x_{1}^{2}+y_{1}^{2}+z_{1}^{2}-r^{2},    
        \label{eq:c}
        \end{equation}
        
        where $\left(x_{1},y_{1},z_{1}\right)$ can be conveniently set to ${\textcolor{violet}{\bullet}} \, \left(x_{\mathrm{sea}},y_{\mathrm{sea}},z_{\mathrm{sea}}\right)$, i.e. the original shower position at sea. There are in general two solutions for $d_{\mathrm{intersect}}$ given by
        \begin{equation}
            d_{\mathrm{intersect}}=\frac{-b\pm\sqrt{b^{2}-4ac}}{2a},
        \label{eq:solutions}
        \end{equation}
        however, only the smaller of the two solutions is of interest, as the larger one is the distance to the intersection point on the other side of the sphere approximating the Earth. Thus, the value of $d_{\mathrm{intersect}}$ can be obtained by inserting Eq.~\ref{eq:a}, \ref{eq:b}, \ref{eq:c}, and $r={\color{teal}R_{\mathrm{Earth}}}+z_{\mathrm{can}}$ into Eq.~\ref{eq:solutions}, and used to compute the intersection point: 
        \begin{equation}
            \left(\begin{array}{c}
            x_{\mathrm{p}}\\
            y_{\mathrm{p}}\\
            z_{\mathrm{p}}
            \end{array}\right)=\left(\begin{array}{c}
            x_{\mathrm{sea}}+c_{x}^{\mathrm{p}}\cdot d_{\mathrm{intersect}}\\
            y_{\mathrm{sea}}+c_{y}^{\mathrm{p}}\cdot d_{\mathrm{intersect}}\\
            z_{\mathrm{sea}}+c_{z}^{\mathrm{p}}\cdot d_{\mathrm{intersect}}
            \end{array}\right).
        \end{equation}
    }
    \item {
    After finding $\textcolor{darkgreen}{\ensuremath{\bullet}} \left(x_{\mathrm{p}},y_{\mathrm{p}},z_{\mathrm{p}}\right)$, the next step is to determine a rotation, which must be applied to move it to $\textcolor{electriclime}{\bullet} 
    \left(x_{\mathrm{can}},y_{\mathrm{can}},z_{\mathrm{can}}\right)$.
    }

    \hfill
    
    \begin{enumerate}
        \item {
        The axis of rotation described by a vector $\vec{a}$ must be identified. If the vectors pointing from the origin of the coordinate system (centre of the Earth) to
        $\textcolor{darkgreen}{\ensuremath{\bullet}} \left(x_{\mathrm{p}},y_{\mathrm{p}},z_{\mathrm{p}}\right)$ and $\textcolor{electriclime}{\bullet} \left(x_{\mathrm{can}},y_{\mathrm{can}},z_{\mathrm{can}}\right)$ are denoted as $\vec{p}$ and $\vec{q}$ respectively, then $\vec{a}$ will be a vector orthogonal to both of them (see Fig.~\ref{fig:axis-of-rotation}), which can be found from the cross-product: 
        \begin{equation}
            \vec{a}=\vec{p}\times\vec{q}=\left[\begin{array}{c}
            y_{\mathrm{p}}z_{\mathrm{can}}-z_{\mathrm{p}}y_{\mathrm{can}}\\
            z_{\mathrm{p}}x_{\mathrm{can}}-x_{\mathrm{p}}z_{\mathrm{can}}\\
            x_{\mathrm{p}}y_{\mathrm{can}}-y_{\mathrm{p}}x_{\mathrm{can}}
            \end{array}\right].
        \end{equation}
        For the rotation, the components of the versor (unit vector) $\hat{a}$ must be computed:
        {\small \begin{equation}
            {
            \hat{a}=\frac{\vec{a}}{\left\Vert \vec{a}\right\Vert }=\left[\begin{array}{c}
            a_{x}\\
            a_{y}\\
            a_{z}
            \end{array}\right]
            .
            }
        \end{equation}}             
        }
        \item {
        Next, the rotation angle ${\color{cyan}\alpha}$ must be determined. It is efficient to directly evaluate its sine and cosine by making use of the following properties of vectors:
        \begin{equation}
            \sin{\color{cyan}\alpha}=\frac{\left\Vert \vec{p}\times\vec{q}\right\Vert }{\left\Vert \vec{p}\right\Vert \cdot\left\Vert \vec{q}\right\Vert }=\frac{\left\Vert \vec{a}\right\Vert }{\left\Vert \vec{p}\right\Vert \cdot\left\Vert \vec{q}\right\Vert }
        \end{equation}
            and
        \begin{equation}
            \cos{\color{cyan}\alpha}=\frac{\left\Vert \vec{p}\cdot\vec{q}\right\Vert }{\left\Vert \vec{p}\right\Vert \cdot\left\Vert \vec{q}\right\Vert }.
        \end{equation}
        }
        \item {
        Finally, the rotation matrix $M$ can be computed, as all the elements are known:
        {\small\begin{equation}
        {
             M=\left[\begin{array}{ccc}
            \cos{\color{cyan}\alpha}+a_{x}^{2}(1-\cos{\color{cyan}\alpha}) & a_{x}a_{y}(1-\cos{\color{cyan}\alpha})-a_{z}\sin{\color{cyan}\alpha} & a_{x}a_{z}(1-\cos{\color{cyan}\alpha})+a_{y}\sin{\color{cyan}\alpha}\\
            a_{y}a_{x}(1-\cos{\color{cyan}\alpha})+a_{z}\sin{\color{cyan}\alpha} & \cos{\color{cyan}\alpha}+a_{y}^{2}(1-\cos{\color{cyan}\alpha}) & a_{y}a_{z}(1-\cos{\color{cyan}\alpha})-a_{x}\sin{\color{cyan}\alpha}\\
            a_{z}a_{x}(1-\cos{\color{cyan}\alpha})-a_{y}\sin{\color{cyan}\alpha} & a_{z}a_{y}(1-\cos{\color{cyan}\alpha})+a_{x}\sin{\color{cyan}\alpha} & \cos{\color{cyan}\alpha}+a_{z}^{2}(1-\cos{\color{cyan}\alpha})
            \end{array}\right].
        }
        \label{eq:Rotation-matrix}
        \end{equation}
        }
        }
        
        A derivation of Eq.~\ref{eq:Rotation-matrix} may be found in Sec.\ 9.2 of \cite{RotationMatrixDerivation}. The matrix $M$ is applied both to position and direction vectors:
        \begin{equation}
            \left[\begin{array}{c}
            x\\
            y\\
            z
            \end{array}\right]_{\mathrm{rotated}}=M\left[\begin{array}{c}
            x\\
            y\\
            z
            \end{array}\right],
        \end{equation}
        \begin{equation}
            \left[\begin{array}{c}
            c_{x}\\
            c_{y}\\
            c_{z}
            \end{array}\right]_{\mathrm{rotated}}=M\left[\begin{array}{c}
            c_{x}\\
            c_{y}\\
            c_{z}
            \end{array}\right].
        \end{equation}
    \end{enumerate}
\end{enumerate}

\subsubsection{Rotation of showers around the Earth}
\label{subsec:rotation-around-Earth}

\hfill

The procedure of reorienting the CORSIKA showers is multi-staged:
\begin{enumerate}
    \item {
    The primary direction is rotated such that the trajectory intersects the middle of the can $\left(0,0,\frac{{\color{orange}h_{\mathrm{can}}}}{2}\right)$ and is computed by
    \begin{equation}
        \left[\begin{array}{c}
        c_{x,}^{\mathrm{mid}}\\
        c_{y}^{\mathrm{mid}}\\
        c_{x}^{\mathrm{mid}}
        \end{array}\right]=M_{\mathrm{mid}}\cdot\left[\begin{array}{c}
        c_{x,}^{\mathrm{sea}}\\
        c_{y}^{\mathrm{sea}}\\
        c_{x}^{\mathrm{sea}}
    \end{array}\right],
    \end{equation}
     where ${\color{orange}h_{\mathrm{can}}}$ is the height of the can and $M_{\mathrm{mid}}$ is computed according to \ref{subsec:Computing-the-needed-rotation}.
    }

    \hfill
    
    \item {
    The surface of the can is increased by \emph{DistaMax} (as described in Sec.\ \ref{sec:weights}) to account for the lateral spread of the muons due to scattering.  
    }

    \hfill
    
    \item {
    A point on the enlarged surface of the can is picked randomly. The code selects whether the point will be on the top cap or the side of the can. The probability of landing on either of the two is weighted by the top and side areas of the can (extended by \emph{DistaMax}) $A_{\mathrm{top}}$ and $A_{\mathrm{side}}$, projected onto the plane perpendicular to $\left[c_{x,}^{\mathrm{mid}},c_{y}^{\mathrm{mid}},c_{z}^{\mathrm{mid}}\right]$.

    \hfill
    
    \begin{enumerate}
        \item {
        If the top area is selected: a random point is drawn from the circle of radius $r_{\mathrm{can}}+\mathrm{\emph{DistaMax}}\cdot c_{x}^{\mathrm{mid}}$.
        }

        \hfill
        
        \item {
        If the side area is selected: a random point is drawn from half the cylinder side of radius $r_{\mathrm{can}}+\mathrm{\emph{DistaMax}}\cdot c_{x}^{\mathrm{mid}}$ and height equal to ${\color{orange}h_{\mathrm{can}}}+\mathrm{\emph{DistaMax}}\cdot\sqrt{1-\left(c_{x}^{\mathrm{mid}}\right)^{2}}$. Only half of the side is used because the shower does not “see” the whole side of the can. This half is subsequently rotated to face the incoming primary.
        }
        
        \hfill
        
    \end{enumerate}
    }
    \item {
    When the point on the increased can surface $\left(x_{\mathrm{can}},y_{\mathrm{can}},z_{\mathrm{can}}\right)$ is selected, the rotation matrix $M$ is computed for that point, following \ref{subsec:Computing-the-needed-rotation}.
    }

    \hfill

    \item {
    The rotation represented by $M$ is applied to the positions and directions of all tracks in an event, regardless of whether they are going to be propagated or not. The final effect is a coherent rotation of the entire shower around the Earth.
    }
\end{enumerate}

\subsubsection{Caveats of shower rotation}
\label{subsec:rotation-bias}

Note that although the rotation described in this section preserves the original shower zenith angle and altitude (and thus the slant depth), it does alter the latitude and/or longitude. This implies a different atmospheric density profile and geomagnetic field strength, both of which are inputs for CORSIKA simulation. The effect is most prominent for inclined showers, as those need to be moved the most to hit the can. It must be stressed that CORSIKA does not allow a variable magnetic field or atmospheric density. This means that one is forced to either accept the inherent larger uncertainty in the case of inclined showers or generate separate simulations for different primary particle zenith and azimuth bins, which in many scenarios will simply be impractical. The impact on the horizontal events will also depend greatly on the geographic location of the detector and the associated variability of the conditions. The impact on the results was tested by running small CORSIKA test simulations for the two locations:
\begin{itemize}
    \item KM3NeT/ARCA ($36.27\degree$N, $16.10\degree$E): $B_x=27.86\,\mathrm{\upmu T},\, B_z=35.39\,\mathrm{\upmu T}$,
    \item KM3NeT/ORCA ($42.80\degree$N, $6.03\degree$E): $B_x=24.25\,\mathrm{\upmu T},\, B_z=39.82\,\mathrm{\upmu T}$,
\end{itemize}
where the magnetic field strength was taken from the IGRF model \cite{IGRF} and a fixed averaged atmospheric density profile was used. The ARCA and ORCA sites are separated by roughly 800$\,$km and the magnetic field components differ by $15\%$ and $12\%$, respectively, leading to a variation of event rates at sea level of the order of $1-10\%$, depending on the primary energy. A variability of $3-16\%$ was also observed by performing the simulation for the two sites with a fixed magnetic field but using two different atmospheric density profiles (obtained with GDAS \cite{GDAS}). The values provided here are only meant to guide the basic intuition on the expected magnitude of such effects; the duty of the users is to perform a dedicated study for their own experimental site.

\subsection{Header}

\hfill

\label{sec:header}

As demonstrated in Listing \ref{lst:header}, the header of the gSeaGen output files contains all the basic information about the simulation. This includes the low- and high-energy hadronic interaction models used by CORSIKA, the code used internally for transport of muons through matter (in this case PROPOSAL), CORSIKA and gSeaGen versions, energy cuts, the spectral index used in the CORSIKA simulation, can position and dimensions (coordinates of the can centre, its height, and radius), depth, site coordinates, cosmic ray flux model, primary particle and the generation spectrum. An example is shown below:

\begin{lstlisting}[
language={python}, 
caption={Header example for a CORSIKA file processed with gSeaGen.},
label={lst:header},
captionpos=t
]
start_run: 1
drawing: surface
physics: CORSIKA UrQMD SIBYLL-2.3d
physics: gSeaGen PROPOSAL 6.1.5
simul: CORSIKA 7.7410 230627 0000
simul: gSeaGen v7.4.0 230930 2019
cut_primary: 1000 1e+06 -0.866 -0.866
cut_seamuon: 999 0 0 0
cut_nu: 999 0 0 0
spectrum: 1
fixedcan: 0 0 0 957 818.4
genvol: 0 0 0 0 1000
depth: 3450
primary: 2212
flux: 2212 GST3 /some/directory/DAT000001
\end{lstlisting}

\section{Code testing and validation}
\label{sec:code_testing}
\hfill

To ensure good code quality and stability, gSeaGen has a continuous integration and continuous deployment (CI/CD) pipeline set up in both its private and \href{https://git.km3net.de/opensource/gseagen/-/pipelines}{public repository}. The pipeline is automatically executed at each commit and is divided into several stages. The first is devoted to the generation and deployment of documentation in the form of internal notes generated from LaTeX and \href{https://opensource.pages.km3net.de/gseagen/}{the Doxygen code documentation}. The next focusses on verifying adherence to the fixed code style (LLVM) using clang-format and code quality with cppcheck and flawfinder. Subsequently, the code compilation is tested, and then a set of benchmark jobs is run to verify whether gSeaGen works properly, testing different modes of neutrino generation with GENIE (v3.4.0) and processing of muons from CORSIKA (v7.7410) using PROPOSAL (v6.1.5). Finally, if the code was tagged, it is put in the docker container, converted to the singularity container, and uploaded to the SFTP server from where it can be retrieved. The source code and the singularity image are also automatically uploaded to \href{https://zenodo.org/records/13989047}{the gSeaGen Zenodo record}. CI/CD has been recently upgraded as a result of the internal code review, carried out within the KM3NeT Collaboration.

With regard to the specific developments presented in this publication, the critical parts of the code were reimplemented in parallel in Python. The main motivation was to verify whether the results can be reproduced and to make visualisation easier. The attempt was successful and most of the presented plots were produced using the Python code, which will be included in the gSeaGen repository in the form of jupyter notebooks.

The new gSeaGen functionality was used to produce several smaller CORSIKA test simulations and a big one with more than $1.44\cdot 10^{10}$ showers generated with primary energies between $1\,\mathrm{TeV}$ and $8\,\mathrm{EeV}$ (more details can be found in \cite{my-Thesis}). The results obtained with gSeaGen were cross-checked against MUPAGE \cite{MUPAGE}, a standard KM3NeT code that generates muons directly on the surface of the can, and a good level of agreement was found. Furthermore, both simulation codes were compared to the experimental data \cite{my-ICRC2021,my-Thesis,Andrey_and_me_ICRC2023}. In the case of the CORSIKA simulation, a clear deficit of muon events was found at high energies, which is consistent with observations of other experiments and is commonly referred to as the Muon Puzzle \cite{MuonPuzzle}.

\section*{Acknowledgements}

\noindent The authors acknowledge the financial support of the funding agencies:
Funds for Scientific Research (FRS-FNRS), Francqui foundation, BAEF foundation.
Czech Science Foundation (GAČR 24-12702S);
Agence Nationale de la Recherche (contract ANR-15-CE31-0020), Centre National de la Recherche Scientifique (CNRS), Commission Europ\'eenne (FEDER fund and Marie Curie Program), LabEx UnivEarthS (ANR-10-LABX-0023 and ANR-18-IDEX-0001), Paris \^Ile-de-France Region, Normandy Region (Alpha, Blue-waves and Neptune), France;
Shota Rustaveli National Science Foundation of Georgia (SRNSFG, FR-22-13708), Georgia;
This work is part of the MuSES project which has received funding from the European Research Council (ERC) under the European Union’s Horizon 2020 Research and Innovation Programme (grant agreement No 101142396).
The General Secretariat of Research and Innovation (GSRI), Greece;
Istituto Nazionale di Fisica Nucleare (INFN) and Ministero dell’Universit{\`a} e della Ricerca (MUR), through PRIN 2022 program (Grant PANTHEON 2022E2J4RK, Next Generation EU) and PON R\&I program (Avviso n. 424 del 28 febbraio 2018, Progetto PACK-PIR01 00021), Italy; IDMAR project Po-Fesr Sicilian Region az. 1.5.1; A. De Benedittis, W. Idrissi Ibnsalih, M. Bendahman, A. Nayerhoda, G. Papalashvili, I. C. Rea, A. Simonelli have been supported by the Italian Ministero dell'Universit{\`a} e della Ricerca (MUR), Progetto CIR01 00021 (Avviso n. 2595 del 24 dicembre 2019); KM3NeT4RR MUR Project National Recovery and Resilience Plan (NRRP), Mission 4 Component 2 Investment 3.1, Funded by the European Union – NextGenerationEU,CUP I57G21000040001, Concession Decree MUR No. n. Prot. 123 del 21/06/2022;
Ministry of Higher Education, Scientific Research and Innovation, Morocco, and the Arab Fund for Economic and Social Development, Kuwait;
Nederlandse organisatie voor Wetenschappelijk Onderzoek (NWO), the Netherlands;
The National Science Centre, Poland, grant number 2021/41/N/ST2/01177;
We gratefully acknowledge the funding support by program “Excellence initiative-research university” for the AGH University in Krakow as well as the ARTIQ project: UMO-2021/01/2/ST6/00004 and ARTIQ/0004/2021;
The grant “AstroCeNT: Particle Astrophysics Science and Technology Centre”, carried out within the International Research Agendas programme of the Foundation for Polish Science financed by the European Union under the European Regional Development Fund;
Ministry of Research, Innovation and Digitalisation, Romania;
Slovak Research and Development Agency under Contract No. APVV-22-0413; Ministry of Education, Research, Development and Youth of the Slovak Republic;
MCIN for PID2021-124591NB-C41, -C42, -C43 and PDC2023-145913-I00 funded by MCIN/AEI/10.13039/501100011033 and by “ERDF A way of making Europe”, for ASFAE/2022/014 and ASFAE/2022 /023 with funding from the EU NextGenerationEU (PRTR-C17.I01) and Generalitat Valenciana, for Grant AST22\_6.2 with funding from Consejer\'{\i}a de Universidad, Investigaci\'on e Innovaci\'on and Gobierno de Espa\~na and European Union - NextGenerationEU, for CSIC-INFRA23013 and for CNS2023-144099, Generalitat Valenciana for CIDEGENT/2018/034, /2019/043, /2020/049, /2021/23, for CIDEIG/2023/20 and for GRISOLIAP/2021/192 and EU for MSC/101025085, Spain;
Khalifa University internal grants (ESIG-2023-008 and RIG-2023-070), United Arab Emirates;
The European Union's Horizon 2020 Research and Innovation Programme (ChETEC-INFRA - Project no. 101008324).

\label{sec:acknowledgements}

\bibliographystyle{elsarticle-num}
\bibliography{main}
\end{document}